\documentclass{aa}  

\usepackage[normalem]{ulem} 
\usepackage{natbib}

\usepackage{graphicx}
\usepackage{xcolor}
\usepackage{xspace}
\usepackage{txfonts}

\usepackage{lscape}             
                                
\usepackage{placeins}           
\usepackage{hyperref}

\newcommand*{\old}[1]{}

\newcommand*{\comment}[1]{}

\newcommand{\be}{\begin{equation}}
\newcommand{\ee}{\end{equation}}
\newcommand{\bea}{\begin{eqnarray}}
\newcommand{\eea}{\end{eqnarray}}

\newcommand{\m}{\,\hbox{m}}
\newcommand{\mm}{\,\hbox{mm}}
\newcommand{\km}{\,\hbox{km}}  
\newcommand{\mum}{\,\text{µm}}
\newcommand{\cm}{\,\hbox{cm}}

\newcommand{\au}{\,\hbox{au}}
\newcommand{\g}{\,\hbox{g}}
\newcommand{\s}{\,\hbox{s}}

\newcommand{\yr}{\,\hbox{yr}}   
\newcommand{\Myr}{\,\hbox{Myr}}

\newcommand{\erg}{\,\hbox{erg}}
\newcommand{\K}{\,\hbox{K}}  
\newcommand{\total}{\,\hbox{d}}

\newcommand{\kms}{\km\s^{-1}}
\newcommand{\ms}{\m\s^{-1}}
\newcommand{\gccm}{\g\cm^{-3}}
\newcommand{\ergg}{\erg\g^{-1}}
\newcommand{\aupx}{\,\hbox{au/pixel}}

\newcommand{\sbs}[1]{_\text{#1}}

\renewcommand{\d}{\ensuremath{\mathrm{d}}}

\newcommand{\QD}{\ensuremath{Q\sbs{D}^*}}
\newcommand{\p}{\sbs{p}}
\renewcommand{\t}{\sbs{t}}
\newcommand{\imp}{\ensuremath{\sbs{imp}}}

\newcommand{\reference}{Reference\xspace }

\newcommand{\zodisteep}{Zodi\xspace }

\newcommand{\fomaref}{Fomalhaut\xspace }

\begin{document}

   \title{How hard is dust in debris disks?}

   \author{Tobias Stein\inst{1}
        \and Alexander V. Krivov\inst{1}
        \and Torsten Löhne\inst{1}
        }

   \institute{Astrophysikalisches Institut und Universit\"ats-Sternwarte, Friedrich-Schiller-Universit\"at Jena, Schilleg\"a{\ss}chen 2-3, 07745 Jena, Germany\\
             \email{tobias.stein@uni-jena.de}
             }

   \date{Received 10 December 2025 / Accepted 09 January 2026}

  \abstract{
  
  Observational appearance of debris disks is largely controlled by collisional grinding of their dust grains. However, the mechanical strength of dust at sizes in the micrometer to millimeter range is poorly known.
  Recent studies suggested that dust particles in the Solar system might have a higher critical fragmentation energy $\QD$ value than previously anticipated. Another recent study considered the Fomalhaut debris disk and found lower $\QD$ values to provide better fits to the data. In order to constrain the mechanical strength of dust, we investigate collisional evolution of debris disks with $\QD$ prescriptions differing by $\sim 3$ orders of magnitude. We find that, above a certain threshold $\QD$ value, the disk's collisional evolution is dominated by rebounding~-- rather than disruptive or cratering~-- collisions. Rebounding (a.k.a. bouncing) collisions are those in which both impactors survive, being only slightly eroded and producing fragments that only carry a minor fraction of their mass. We show that disks dominated by rebounding collisions would have brightness profiles increasing outward outside the parent belt. Since such profiles are not observed, this places an upper limit on how hard the debris dust is allowed to be in order not to violate the observations. We derive an approximate analytic expression for this limit: $\QD \approx (1/8) v\sbs{K}^2(r)$ for grains close to the radiation pressure blowout size, where $v\sbs{K}$ in the Keplerian circular speed at a distance $r$ from the star. This implies $\QD \lesssim 10^{9...10} \ergg$ for micrometer-sized grains in typical debris disks. Even though rebounding collisions are not expected to affect debris disk evolution significantly, we emphasize that these collisions are actually much more frequent than disruptive and cratering ones in all debris disks.
  }


   \maketitle

\nolinenumbers
\section{Introduction}

Debris disks encircling a sizeable fraction of stars are collections of solid bodies spanning a broad size range, perhaps from dwarf planets as large as $\sim 1000\km$ down to micrometer-sized dust \citep[see, e.g.][for reviews]{wyatt-2008,krivov-2010,matthews-et-al-2013,wyatt-2020,pearce-2024}. However, what is only observed directly is thermal emission of, or stellar light scattered by, dust. Thus, to interpret debris disk observations, one needs to understand the production and sustainment of dust in those disks.

As the very name ``debris disk'' readily suggests, the key process is destructive collisions. The cascade of collisions grinds larger bodies into ever-smaller ones, down to dust grains. How these collisions exactly work at dust sizes, and thus how the disks appear in observations, depend crucially on the relative velocities of disk particles and their tensile strength. While the former can be estimated, e.g., from the vertical thickness of the disk \citep[e.g.,][]{matra-et-al-2019b,zawadzki-et-al-2026}, little is known about the latter. Laboratory impact experiments at the (relatively high) velocities and (relatively small) sizes in question \citep[e.g.,][]{fujiwara-et-al-1977,davis-ryan-1990,housen-holsapple-1999,nakamura-et-al-2015,landeck-2023,horikawa-et-al-2025} are challenging, and numerical simulations \citep[e.g.,][and references therein]{benz-asphaug-1999,leinhardt-stewart-2012} do not necessarily yield reliable results either. Both are hampered by the poor knowledge of the material composition of debris dust grains and their morphology (for example, possible microporosity).

Recent studies suggest that debris dust could be more difficult to destroy than is usually assumed. \citet{rigley-wyatt-2022} and, more recently, \citet{pokorny-et-al-2024} fitted zodiacal cloud observations in the Solar system with their collisional and dynamical models to conclude that interplanetary dust in the Solar system must be by 2-3 orders of magnitude more resistant to collisions than previously thought. \citet{sommer-et-al-2025}, in contrast, found that the weaker dust works best to explain the extended sheet of warm dust discovered by {\sl JWST} in the Fomalhaut system, interior to the main disk \citep{gaspar-et-al-2023}.

A standard quantity to characterize the collisional strength of solids is their critical fragmentation energy $\QD$, which is the minimum impact energy per unit mass required to disrupt a body. The idea of this paper is to explore several $\QD$ models, including a commonly used model to characterize monolithic grains \citep{krivov-et-al-2018}, an alternative stronger model presented in \citet{rigley-wyatt-2022}, as well as a weak one extrapolated from \citet{sommer-et-al-2025}. For each $\QD$ model, we run a collisional code to determine the resulting size and radial distributions of dust in a typical disk. We then simulate typical observables, such as spectral energy distributions and radial profiles of brightness, for these models. The goal is to see which $\QD$ models make sense and which clearly contradict the observations.

An additional aspect of this work is collisional outcomes. If, for instance, the dust is so strong that the typical impact energy is not sufficient to destroy the colliders of some size, it does not necessarily mean that collisions do nothing. Milder outcomes, such as cratering of one or both colliders, bouncing, or even sticking and growth, might still be possible. Our collisional modeling does include a realistic description of all these collisional outcomes. Therefore, in this study, we also explore which types of collisions are important and how they may affect the resulting distributions of dust and the observational appearance of debris disks.

In Section~\ref{Sec:QD}, we describe the $\QD$ models which will be compared in this work,  Section~\ref{Sec:outcomes} characterizes possible collisional outcomes we consider and Section~\ref{Sec:code} describes our collisional code and simulation setup. We present the results in Section~\ref{Sec:Results}, discuss them in Section~\ref{Sec:Discussion} and draw our conclusions in Section~\ref{Sec:Conclusion}.

\section{Critical fragmentation energy}
\label{Sec:QD}

\subsection{General prescription}

The outcomes of collisions between the objects sensitively depend on the
critical fragmentation energy, which is defined as the minimum impact energy needed
to disrupt the body, per unit mass. It depends on 
the object radius $s$ and the relative (or impact, or collisional) velocity $v_{\text{rel}}$.
Smaller bodies are kept together by molecular forces
(``strength regime''), whereas larger objects are bound by gravity (``gravity regime'').
Accordingly, a standard prescription is a sum of two power laws, one for each regime \citep{benz-asphaug-1999}:
\be
\QD=
A_\mathrm{s}\left(s \over 1\cm \right)^{b_\mathrm{s}}
+
A_\mathrm{g}\left(s \over 1\cm\right)^{b_\mathrm{g}} ,
\label{eq:QD1}
\ee
where coefficients $A_\mathrm{s}$ and $A_\mathrm{g}$ depend on the relative speed $v_\mathrm{rel}$:
\be
\QD=
\left[
A_\mathrm{s}^0\left(s \over 1\cm \right)^{b_\mathrm{s}}
+
A_\mathrm{g}^0\left(s \over 1\cm\right)^{b_\mathrm{g}}
\right]
\left(
v_\mathrm{rel} \over v_0
\right)^{b_\mathrm{v}} .
\label{eq:QD2}
\ee
In what follows, we set $v_0 = 3\km\s^{-1}$ and $b_\mathrm{v}=0.5$,
following \citet{stewart-leinhardt-2009}.

\subsection{Typical relative velocities in debris disks}
\label{typical_relative_velocities}

The typical collisional velocities can be inferred from the measured geometric vertical thickness of the disks at (sub)mm wavelengths, assuming equipartition of energy \citep[see][however]{jankovic-et-al-2024}.
For instance, \citet{terrill-et-al-2023} analyzed 16 highly inclined debris disks
observed by ALMA. They found a wide range of vertical aspect ratios,
$h$, ranging from $0.020 \pm 0.002$ (AU Mic) to $0.20 \pm 0.03$ (HD 110058).
The average value for disks with moderate fractional width is $h \sim 0.03$
(for wide disks it appears to be somewhat higher, $\sim 0.05$).
Similarly, the REASONS survey of 74 disks \citep{matra-et-al-2025} derived the aspect ratios in the range from
$\sim 0.01$ to $0.3$ for a subset of disks for which the vertical thickness could be constrained. Finally, for a set of 13 most highly-inclined targets observed with ALMA in the ARKS program with the highest resolution achieved today \citep{marino-et-al-2026},
the (HWHM) vertical aspect ratios were found to lie in the range $0.0026 \le h \le 0.193$, with a median best-fit value of $h = 0.021$ \citep{zawadzki-et-al-2026}.
Assuming roughly $h \sim 0.03$ to be a typical value,
eq.~13 of
\citet{zawadzki-et-al-2026} then gives the typical relative velocity in a disk, assuming equipartition between out-of-plane and in-plane velocities:
\be
v_\mathrm{rel} = 2\sqrt{3} h v_\mathrm{K} \approx 0.1 v_\mathrm{K},
\ee
where $v_\mathrm{K}$ is the circular Keplerian speed at the radius of the disk. For our reference disk
with a radius of 100~au around a solar-type star, we get
$v_\mathrm{K} = 3 \km\s^{-1}$ and  $v_\mathrm{rel} \approx 300 \m\s^{-1}$. 

\subsection{Traditional model for $\QD$}

As a reference case (hereafter called ``\reference model''), we use a $\QD$ prescription appropriate for ``monolithic'', collisionally strong planetesimals
taken from \citet[][their eq.~1]{krivov-et-al-2018}:
$A_\mathrm{s}^0=2.7 \times 10^7 \erg\g^{-1}$,
$A_\mathrm{g}^0=0.63\erg\g^{-1}$,
$b_\mathrm{s}=-0.37$,
and $b_\mathrm{g}=1.38$.
This prescription is essentially based on the SPH simulations by \citet{benz-asphaug-1999}
for basalt ($A_\mathrm{s}^0=3.5 \times 10^7 \erg\g^{-1}$,
$A_\mathrm{g}^0=0.81\erg\g^{-1}$,
$b_\mathrm{s}=-0.38$,
and $b_\mathrm{g}=1.36$)
and ice ($A_\mathrm{s}^0=1.6 \times 10^7 \erg\g^{-1}$,
$A_\mathrm{g}^0=1.1\erg\g^{-1}$,
$b_\mathrm{s}=-0.39$,
and $b_\mathrm{g}=1.26$),
complemented with the velocity dependence taken from \citet{stewart-leinhardt-2009}.
It is also close to results by \citet{jutzi-et-al-2010} for basalt, which are
$A_\mathrm{s}^0=2.8 \times 10^7 \erg\g^{-1}$,
$A_\mathrm{g}^0=1.1\erg\g^{-1}$,
$b_\mathrm{s}=-0.38$,
and $b_\mathrm{g}=1.36$.

   \begin{figure*}[htb!]
   \resizebox{\hsize}{!}
        {
        \includegraphics[width=0.49\linewidth]{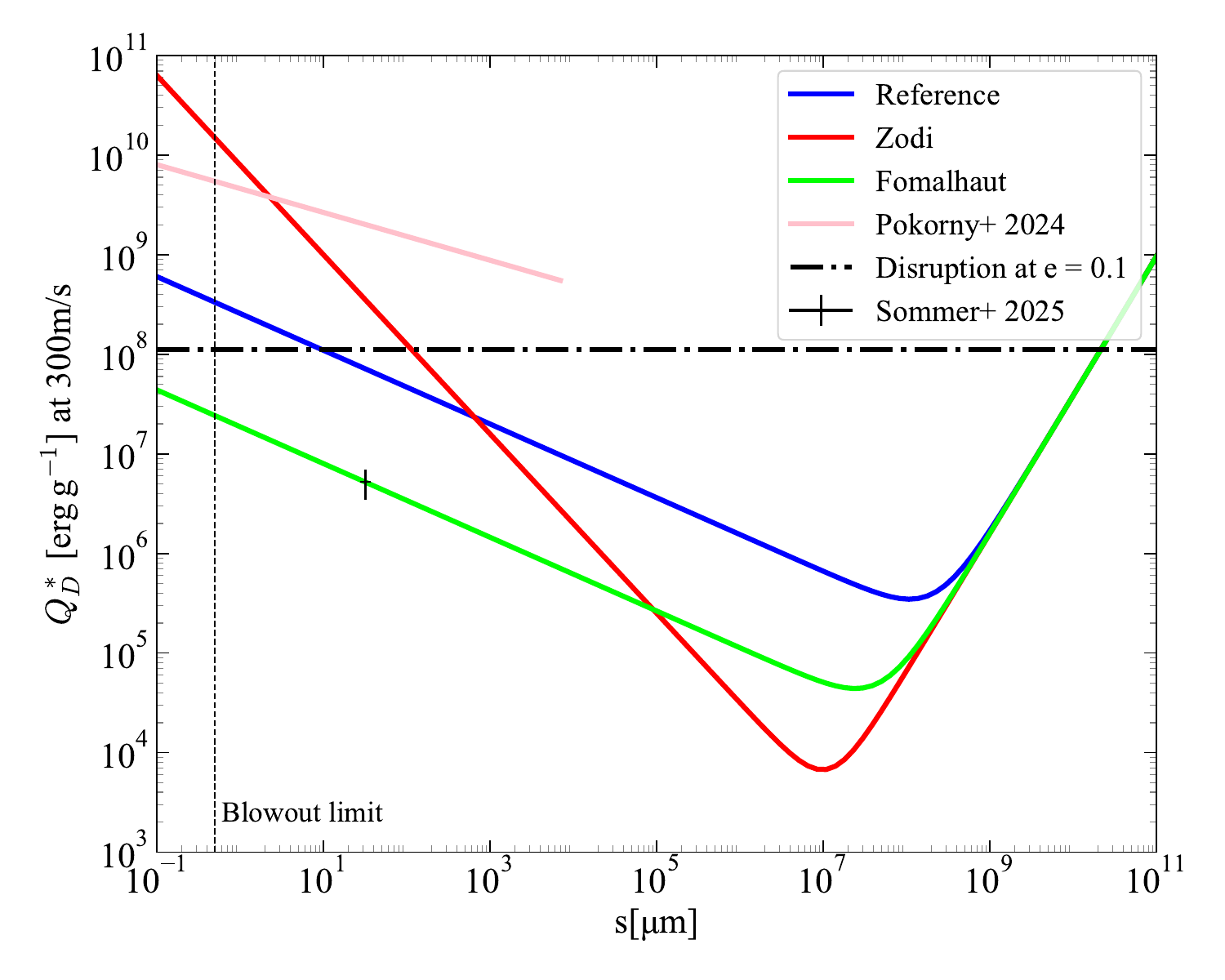}
        \includegraphics[width=0.49\linewidth]{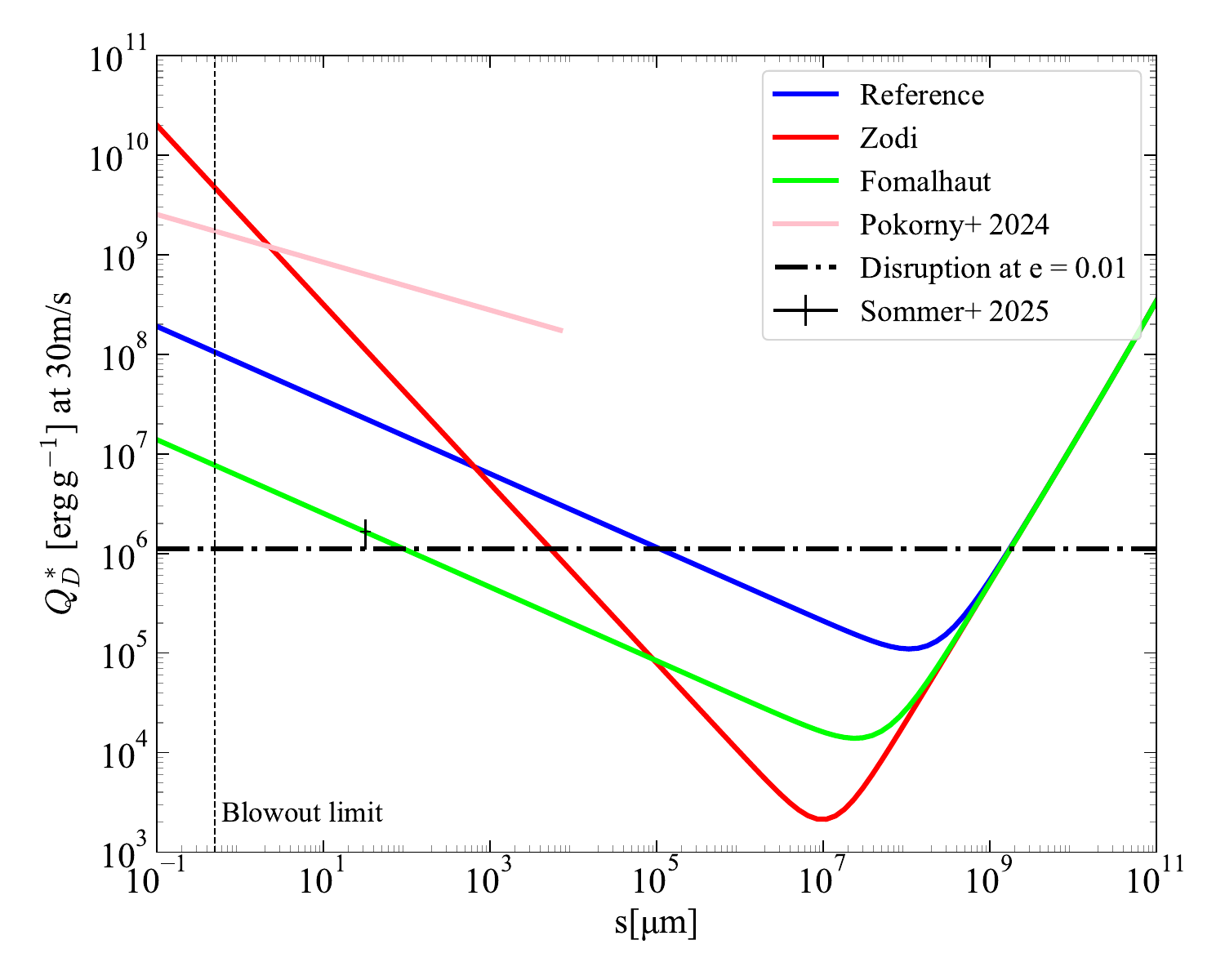}
        }
        \caption{$\QD$ models chosen:
        \reference (blue), \zodisteep (red), and \fomaref (green).
        These are shown for two typical collisional velocities:
        $300\m\s^{-1}$ (left) and $30\m\s^{-1}$ (right).
        For comparison, the $\QD$ fit from \citet{pokorny-et-al-2024} (pink) as well as the \citet{sommer-et-al-2025} constraint at $32\mum$
        (black dot with the uncertainty bar) are plotted as well.
        Horizontal black lines mark an approximate disruption threshold, for the respective velocities. Note that the \citet{pokorny-et-al-2024} model only describes particles with sizes up to $1\cm$.
        }
        \label{fig:QD models}
    \end{figure*}

\subsection{Zodi-based model for $\QD$}

As an alternative case, we define the ``\zodisteep model''.
It is based on the $\QD$ constraints from \citet{rigley-wyatt-2022}. They fitted zodiacal cloud observables with their collisional and dynamical model and find $A_\mathrm{s}=2.0 \times 10^7 \erg\g^{-1}$ and $b_\mathrm{s}=-0.90$
(see their Sections 4.2, 6.5 and Fig.~25).

Note that they give $A_\mathrm{s}$ rather than $A_\mathrm{s}^0$.
To compute the latter, we shall roughly estimate the typical collisional speeds
in their zodi modeling.
Their Fig. 14 gives a distribution of mass input into the dust production
in terms of pericentric distance $q$ and eccentricity $e$. To get a rough idea of the
collisional speed, we consider two intersecting orbits with the same $q$ and $e$.
The relative velocity at the intersection points is maximum when the two orbits are apsidally anti-aligned,
in which case \citep[][their eq.~A10]{costa-et-al-2024}
\be
 v_\mathrm{rel}
 =
 {2e \over \sqrt{1-e} \; (1-e)}
 \sqrt{1\au \over q} \; v_\mathrm{K}(1\au).
\label{eq:v_rel_max}
\ee
The highest mass input in the model of \citet{rigley-wyatt-2022} occurs in a stripe on the
$(q,e)$-plane that roughly corresponds to aphelion distances of $4...5\au$.
This stripe extends from $q\approx 4\au$ and $e\approx 0$ to $q\approx 0.5\au$ and $e\approx 0.9$, with higher values toward lower perihelia (and higher
eccentricities). Equation \ref{eq:v_rel_max} gives a broad range of relative velocities
from $\approx 3 \km\s^{-1}$ to $\approx 130 \km\s^{-1}$~--- again, with preference toward higher values.
As a rough proxy for the entire cloud, we will take $v_\mathrm{rel} \approx 30\km\s^{-1}$.
This gives $A_\mathrm{s}^0=6.3 \times 10^6 \erg\g^{-1}$.

Another model comes from \citet{pokorny-et-al-2024}.
They did a similar analysis for the zodiacal
cloud and found
$A_\mathrm{s}=(5.0 \pm 4.0) \times 10^9 \erg\g^{-1}$  and $b_\mathrm{s}=-0.24$.
Assuming, again, $v_\mathrm{rel} \approx 30\km\s^{-1}$ to be typical of the zodiacal cloud,
this translates to
$A_\mathrm{s}^0=(1.6 \pm 1.3) \times 10^9 \erg\g^{-1}$. However, for simplicity, we will only consider one Zodi-based model and not investigate this model further.

\subsection{Fomalhaut-based model for $\QD$}

Our third model is the ``\fomaref'' one. It is inspired by
\citet{sommer-et-al-2025} who investigated the Fomalhaut debris disk and determined a best fit for the $\QD$ value of $32\mum$-sized particles to be $\QD=(3\pm1)\times10^{6}\ergg$.
That best fit also includes a $\QD$ slope of $b_s = -0.15$ and a maximum value of $b_{s,\text{max}}=-0.45$, but the authors do not place high confidence in this value. This steepness is much closer to our \reference model than to the \zodisteep model. So, to keep things simple, we will extrapolate their result using the slope of the \reference model. While \citet{sommer-et-al-2025} report average collisional velocities of $100\ms$, we will scale their result to
$v_0 = 3\km\s^{-1}$ to calculate $A_\mathrm{s}^0$.

We also note that \citet{rigley-wyatt-2022} and \citet{sommer-et-al-2025} assume a material composition and density different from ours
(astrosilicate, $3.3\g\cm^{-3}$, see Section~\ref{subsec: setup}).
We ignore these differences for two reasons.
First, we wish to clearly see the role of changing the $\QD$, and not several parameters simultaneously.
Second, the expected corrections would only be minor,
compared to the other uncertainties involved.

\subsection{Summary of $\QD$ models}

For simulations in this paper, we thus consider three models:
\reference, \zodisteep, and \fomaref ones, as defined above. To the latter two, we added the term for the gravity regime with the same parameters $A_\mathrm{g}^0=0.63\erg\g^{-1}$
and $b_\mathrm{g}=1.38$ as in the \reference model.

\begin{table}[htb!]
    \centering
    \caption{Summary of simulation parameters used.}
    \begin{tabular}{c|ccccc}
         Run Name & $A_s^0\,\left[10^7\,\frac{\erg}{\g}\right]$ & $b_s$ & $A_g^0\,\left[\frac{\erg}{\g}\right]$ & $b_g$ \\\hline
        \reference & 2.70 & -0.37 & 0.63 & 1.38 \\
        \zodisteep & 0.63 & -0.90 & 0.63 & 1.38\\
        \fomaref & 0.196 & -0.37 & 0.63 & 1.38\\\hline
    \end{tabular}
    \label{tab:setup}
\end{table}

The parameters of all three models are collected in Table~\ref{tab:setup}. The models are depicted in Fig.~\ref{fig:QD models} for different impact speeds of
$300\m\s^{-1}$ (left) and $30\m\s^{-1}$ (right),
corresponding to eccentricity dispersion of 0.1 and 0.01, respectively. Additionally, the model from \citet{pokorny-et-al-2024} is shown for comparison.
It is seen that dust
(i.e., particles up to $\sim 1\mm$ in size) is the weakest in the \fomaref model, harder to disrupt in the \reference one, and the strongest in the \zodisteep model. 
For example, in the \zodisteep model, the smallest ($\mum$-sized) dust grains are 20--30 times harder than in the \reference one.

Overplotted with horizontal lines is the collisional disruption threshold of $\QD$ for collisions of two equal-sized particles, estimated as $v_\mathrm{rel}^2/8$ \citep[see eq.~(5.2) in][]{krivov-et-al-2005}. It shows at which sizes the particles are expected to undergo
disruptive collisions and, conversely, which grains are likely to remain essentially ``indestructible'', being
subject to cratering or rebounding collisions only. Consider, for instance,  the left panel of Fig.~\ref{fig:QD models}
for $300\m\s^{-1}$. 
We see that in the \fomaref model, the material is prone to catastrophic collisions
at all sizes, and in the \reference one all grains larger than $\approx 10\mum$ can be shattered. In contrast,
in the \zodisteep model, grains smaller than hundreds of $\mum$ should become indestructible.

\section{Collisional outcomes}

\label{Sec:outcomes}

\begin{figure*}
    \centering
    \includegraphics[width=\linewidth]{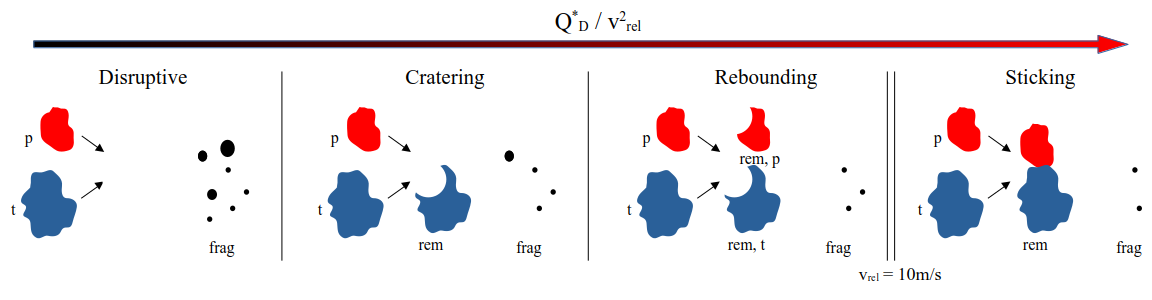}
    \caption{
    Schematic of the four possible collisional outcomes, eqs.~(\ref{eq:disrupt})--(\ref{eq:stick}). The ``projectile'' particle (the less massive of the two colliders, marked with ``p'' and shown in red),  and the ``target'' one (the more massive one, ``t'', in blue) collide to produce one or two remnants (``rem'' or ``rem, p'' and ``rem, t'', each having the color of its progenitor particle) and
    a cloud of fragments that may originate from both colliders (``frag'', shown in black).
    }
    \label{fig:CollisionSchematic}
\end{figure*}

Once $\QD$ and $v_\text{rel}$ are known, the outcome of a collision between two arbitrary objects, the projectile of mass $m\p$ and the target of mass $m\t$ (such that $m\p \le m\t$) is to be determined. The impact energy sets the individual fate of either collider according to the following algorithm \citep{tl-habil}.

First, it is checked whether half the impact energy suffices to disrupt the projectile,
i.e.~whether $E\imp/2 > \QD(m\p) m\p$.
If this is the case, the projectile mass is added to the fragment mass budget and the
remaining energy of the projectile's half is assumed to be imparted on the target.
Otherwise, the projectile is assumed to be cratered but rebounding from the target or sticking to it -- depending on whether the impact velocity
is below or above a threshold of 10~m/s. Owing to this low threshold, sticking collisions do not occur in our simulations considered here.

Second, the remaining energy~-- at least half of the total impact energy -- can either disrupt or crater
the target and lead to the dispersal of either the total mass or part of it, respectively.
This ``equipartition'' of impact energy between the two colliders can only be assumed if their material
strengths are comparable.
What comes out is either the sum of the two remnants (which can be zero if both are dispersed) or two individual remnants, and the
joint fragment mass (which can be as large as the sum of both colliders).

Hence, there are four different combinations for the mass of the remnant(s), $m\sbs{rem}$,
and that of the fragments, $m\sbs{frag}$ (see Fig.~\ref{fig:CollisionSchematic}).
For disruptive (or destructive, or shattering, or catastrophic) collisions, in which both colliders are destroyed, we have
\begin{equation}
  m\sbs{rem} = 0 \quad\mbox{and}\quad
  m\sbs{frag} = m\p + m\t.
  \label{eq:disrupt}
\end{equation}
Cratering (or erosive) collisions are such that the target gets cratered, while the projectile is destroyed:
\begin{eqnarray}
  m\sbs{rem} &=& m\t - m\sbs{frag}(m\t,E\imp - m\p\QD(m\p)),\\
  m\sbs{frag} &=& m\p + m\sbs{frag}(m\t,E\imp - m\p\QD(m\p)).
\end{eqnarray}
Rebounding (or bouncing, or ``double-cratering'') collisions imply that both impactors survive, only undergoing some erosion, and bounce off each other:
\begin{eqnarray}
  m\sbs{rem,p} &=& m\p - m\sbs{frag}(m\p,E\imp/2), \label{eq:reb1}\\ 
  m\sbs{rem,t} &=& m\t - m\sbs{frag}(m\t,E\imp/2), \label{eq:reb2}\\ 
  m\sbs{frag}~~&=& m\sbs{frag}(m\p,E\imp/2) + m\sbs{frag}(m\t,E\imp/2).
  \label{eq:reb3}
\end{eqnarray}
Finally, sticking (or agglomerating) collisions are those in which both colliders essentially merge, with only a minor fraction of their mass going into fragments; these are characterized by
\begin{eqnarray}
  m\sbs{rem} &=& m\p + m\t - m\sbs{frag}(m\p,E\imp/2) - m\sbs{frag}(m\t,E\imp/2),\\
  m\sbs{frag} &=& m\sbs{frag}(m\p,E\imp/2) + m\sbs{frag}(m\t,E\imp/2).
\label{eq:stick}
\end{eqnarray}

\section{Collisional simulations}
\label{Sec:code}

\subsection{The ACE code}

To model the collisional evolution of the disks,
we use the collisional code ACE \citep[``Analysis of Collisional Evolution'';][]
{krivov-et-al-2005,krivov-et-al-2006,loehne-et-al-2007,loehne-et-al-2011,krivov-et-al-2013,loehne-et-al-2017,sende-loehne-2019}.
The code assumes a certain central star and a planetesimal disk of certain size, mass, excitation, and
other parameters such as mechanical and optical properties of disk particles (including \QD) as input. The code simulates the collisional evolution of the debris disk by solving a kinetic Boltzmann-Smoluchowski equation in a 4-dimensional grid which comprises masses of solids, periastron distances, eccentricities, and longitudes of the periastron as phase space variables. The vertical dimension (defined by the orbital inclination and the longitude of the ascending node) is averaged over. Collisional outcomes are treated as described in Section~\ref{Sec:outcomes}. For every instant of time, the distribution of the phase variables is converted into the coupled radial and size distributions of solids~--- from planetesimals to dust.

An auxiliary program, called PHSimage, takes the ACE output as input and computes the observables (both in thermal emission of dust and scattered stellar light). Specifically, in the subsequent sections we compute the time evolution of disk brightness, spectral energy distributions, as well as brightness profiles of our model disks at selected wavelengths from near-infrared to millimeter.

\subsection{Simulation setup}
\label{subsec: setup}
We perform a set of three runs, corresponding to $\QD$ prescriptions listed in Table~\ref{tab:setup}. For all runs, we use a solar-mass star $M_* = M_\odot$ with solar luminosity $L_* = L_\odot$. It is surrounded by an initially narrow and thin debris disk between $95\au$ and $105\au$ with a small semi-opening angle $\epsilon = 0.05$~rad populated initially by particles in the size range $1\mum  \le s\le200\km$, whose orbits have eccentricities uniformly distributed over $0 \le e \le 0.1$. The initial size distribution is given by $\d N \propto s^{-q}\d s$ with $q=(21+b_s)/(6+b_s)$, where $b_s$ is the exponent for the strength-regime part of the $\QD$ prescription \citep{o'brien-greenberg-2003}.
The initial disk mass $M_0$ in all runs is set to $M_0=100\,M_\oplus$, although the runs will be scaled to arrive at the same dust mass at the given instant of time (see Section~\ref{subsec: longterm}).
The longitude of the periapsis is assumed to be distributed uniformly, so that the disk is rotationally symmetric at all times. We disregard collisions involving particles on unbound orbits, stellar wind effects, and Poynting--Robertson drag and do not include gas in the debris disk.
Gravitational interactions between disk solids are not modeled either. We include stellar radiation pressure, 
assuming astrosilicate particles \citep{draine-2003a}
with a density of $\rho_\text{dust}=3.3\gccm$.
To this end, we use the Mie theory \citep{bohren-huffman-1983} to compute the radiation pressure-to-gravity ratio, $\beta$, as a function of grain radius $s$ \citep{burns-et-al-1979}. For this material and the Sun-like central star, the blowout size is $s_\text{blow} = s(\beta=0.5) = 0.5\mum$.

\section{Results}
\label{Sec:Results}

\subsection{Long-term evolution}
\label{subsec: longterm}

\begin{figure}[htb!]
    \centering
    \includegraphics[width=\linewidth]{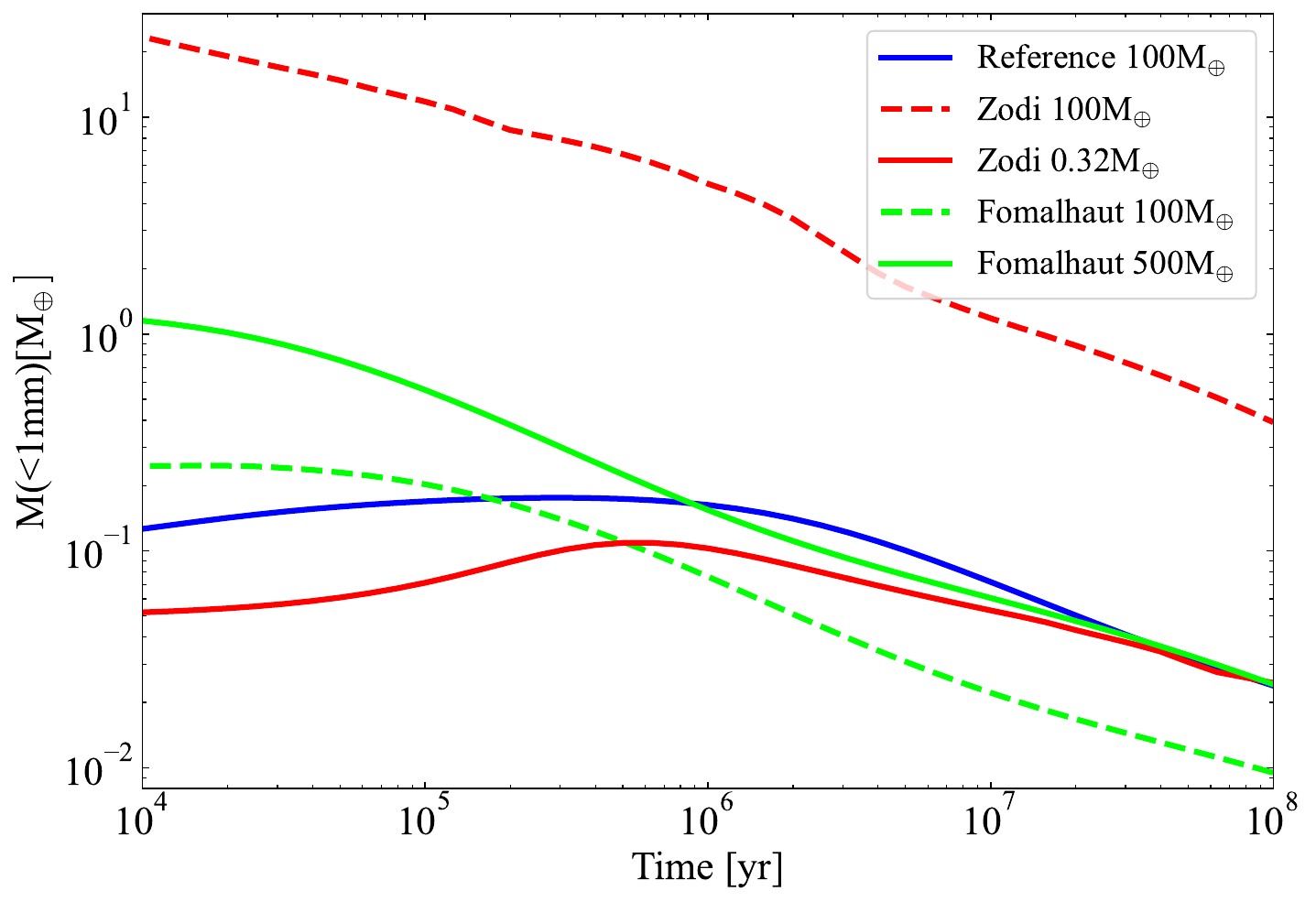}
    \caption{
    Dust mass evolution for the three runs. Dashed lines depict the original (unscaled) and solid lines the scaled versions of the runs. Only one curve is shown for the \reference run as the scaled and unscaled versions are equivalent here.
    }
    \label{fig:Mdust}
\end{figure}

Fig.~\ref{fig:Mdust} (dashed lines) shows the typical temporal evolution of dust in the simulated disks. Although the initial disk mass was set to the same value,  different tensile strengths of material in our runs cause their disks to lose mass at different rates. As a result, the disks contain different amounts of dust, and are unequally bright, at any time instant of interest. Therefore,
for a better comparison of the runs, it is natural to rescale them to have the same amount of dust (or brightness) at some moment of their evolution. 

Such a re-scaling is easily done without re-running the collisional code, by applying the following rule \citep{loehne-et-al-2007,krivov-et-al-2008}.
In a collisionally dominated debris disk in a quasi-steady state with initial disk mass $M_0$ and disk age $t$, for any quantity $F$ that is directly proportional to the disk mass in any size interval in which particles are created and destroyed by collisions (e.g., dust mass), there is a relation
\begin{equation}
    F(xM_0,t/x) = xF(M_0,t),
    \label{shiftLaw}
\end{equation}
valid for any real number $x>0$. However, in any size interval occupied by unbound grains, which are not lost in collisions but get blown out by radiation pressure on (short) dynamical timescales, the following relation holds instead (see Appendix~\ref{app:mass-scaling}):
\begin{equation}
    F(xM_0,t/x) = x^2F(M_0,t).
    \label{shiftLawBeta}
\end{equation}

Specifically, we choose to re-scale the \fomaref and the \zodisteep runs to have the same dust mass (in particles $s<1\mm$) at an arbitrary time instant (which we take to be $100\Myr$) as the \reference run.
Applying eqs.~(\ref{shiftLaw}) and~(\ref{shiftLawBeta}) to the \zodisteep and \fomaref runs, we found $x = 0.0032$ and $x=5.0$, respectively.
The resulting curves are depicted in Fig.~\ref{fig:Mdust} with solid lines. It is these re-scaled runs that are analyzed in the rest of the paper.

\begin{figure}
    \centering
    \includegraphics[width=\linewidth]{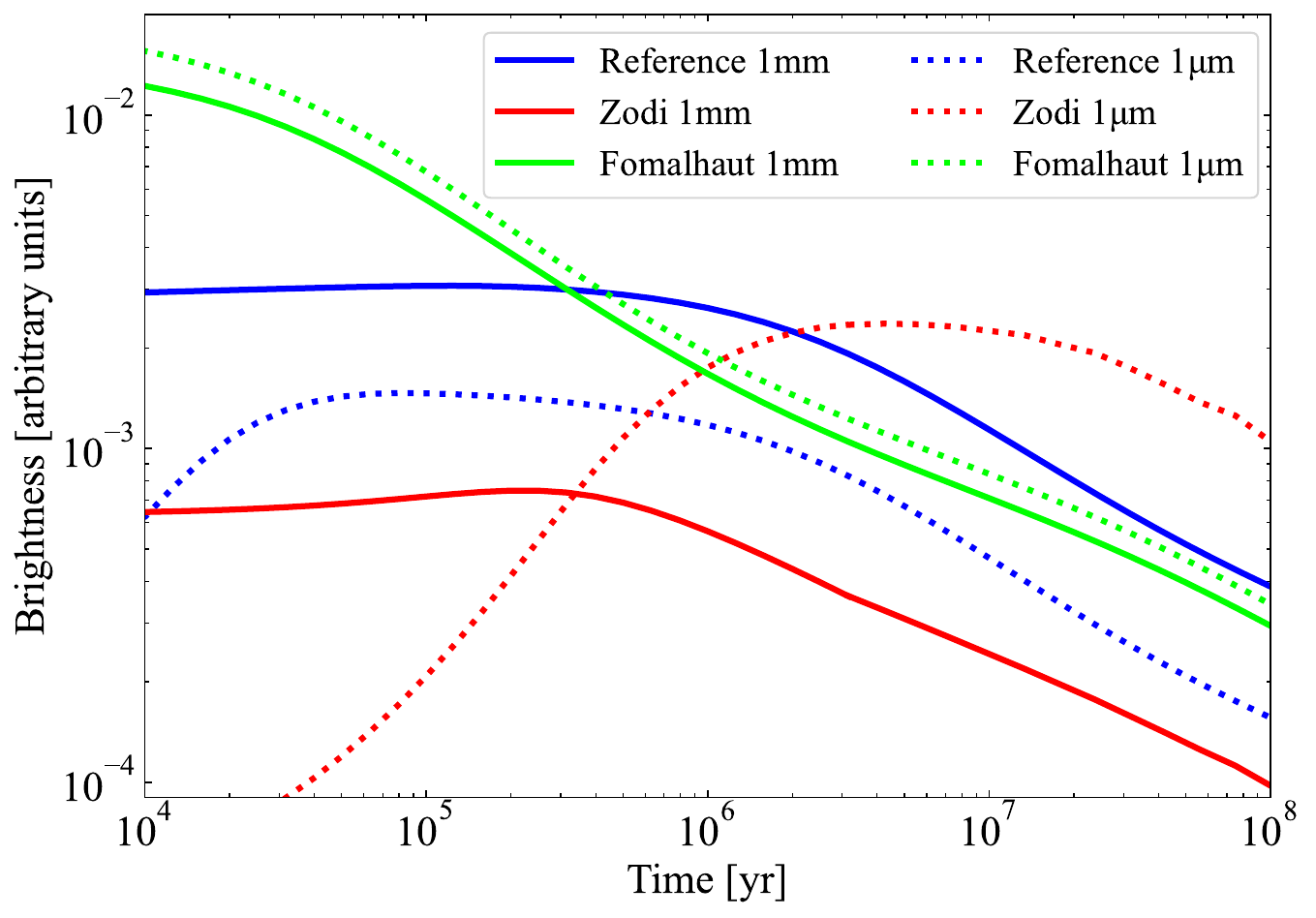}
    \caption{Time evolution of disk brightness in all three runs at $1\mum$ (dashed lines) and $1\mm$ (solid lines).
    }
    \label{fig:brightnessEvolution}
\end{figure}

The dust mass starts to decrease as soon as the collisional lifetime of the weakest particles, called $\tau_b$ in \citet{loehne-et-al-2007}, is reached.
In all three runs, this happens  pretty early, in $\sim 10^4-10^6\yr$. The subsequent loss of dust mass occurs roughly as $\propto t^{-0.5}$, broadly consistent with the theoretical predictions \citep{loehne-et-al-2007} and statistics of debris disks of different ages \citep[e.g.,][]{zuckerman-becklin-1993,habing-et-al-1999,habing-et-al-2001,spangler-et-al-2001, greaves-wyatt-2003,rieke-et-al-2005,moor-et-al-2006,eiroa-et-al-2013,chen-et-al-2014, moor-et-al-2014,holland-et-al-2017,sibthorpe-et-al-2018}. Since the falloff of dust mass is similar in all three models, it is not possible to constrain the mechanical strength of debris dust from the analysis of the long-term evolution.

We have also computed the time evolution of disk brightness at two wavelengths, $1\mm$ and $1\mum$.
These choices correspond to observations with instruments such as SPHERE \citep{beuzit-et-al-2019} or GPI \citep{macintosh-et-al-2014b} and ALMA \citep{wootten-thompson-2009}, respectively. The $1\mum$ profiles are almost entirely made up of scattered stellar light and the $1\mm$ ones represent thermal radiation of the dust itself. As the stellar spectrum, we took the solar photospheric model from \citet{woods-et-al-2009}, and the scattered light was computed with a Mie phase function \citep{bohren-huffman-1983}.

The results are shown in Fig.~\ref{fig:brightnessEvolution}. Not surprisingly, the millimeter flux behaves in a similar way to the dust mass. Interestingly, though, the same applies to the total intensity in scattered light. Note that for the \zodisteep run, the decay of the $1 \mum$ flux sets in later, at several $\Myr$; this is caused by the peculiar behavior of $\mum$-sized grains discussed in detail in Section~\ref{spatialDistribution}, and by the longer time they need to find a steady state. However, the debris disk samples used in statistical studies cover systems older than $\sim 10\Myr$. Thus the conclusion remains the same: one cannot distinguish between our models based on the statistics of disk fluxes at different ages.

\subsection{Size distribution}

As in our previous papers \citep[e.g.][]{krivov-et-al-2006, krivov-et-al-2018}, size distributions in this work will be given as cross-section density distributions per size decade.
That is, we plot the total cross-section area of the disk objects per unit log size bin per unit disk volume. 

\begin{figure}[t!]
    \centering
    \includegraphics[width=1.0\linewidth]{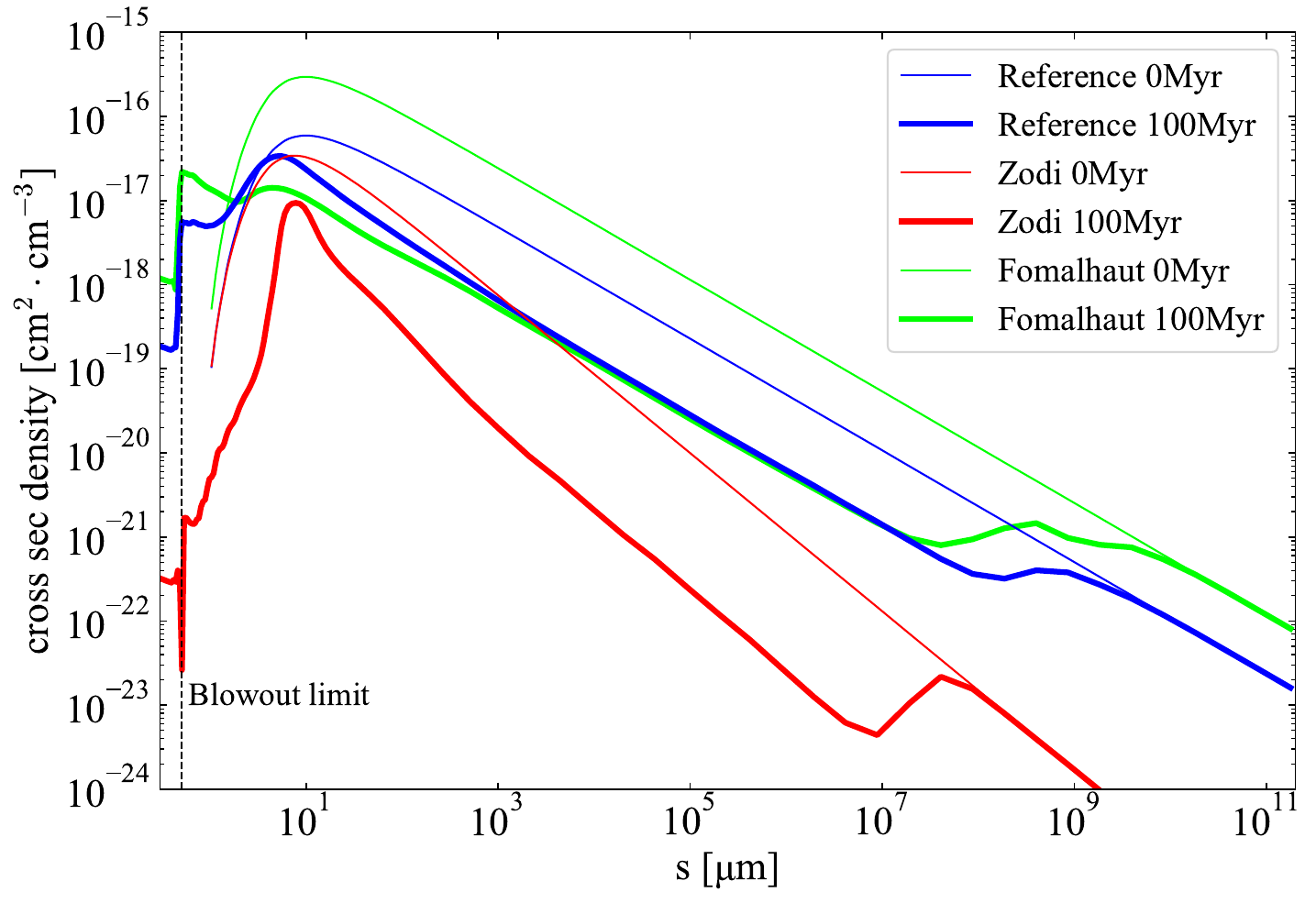}
    \caption{Size distribution at $100\au$ in the three runs in the initial state (thin lines) and at $T=100\,$Myr (thick lines).
    }
    \label{fig:SizeDist}
\end{figure}

The size distributions of the three runs at $100\au$ are plotted in Fig.~\ref{fig:SizeDist}. All three size distributions are typical of collisionally evolved disks. The cross-section density peaks somewhat above the blowout limit: around $7\mum$ for the \zodisteep run, at $5\mum$ for the \reference run, and at $0.5\mum$ for the \fomaref run \cite[see][for a discussion of how the peak position is related to the blowout limit]{pawellek-krivov-2015}. The size distribution to the left of the blowout limit is dominated by dust on hyperbolic orbits.
To the right of the peak, the slope of the distribution is set by the $\QD$ prescription used \citep{o'brien-greenberg-2003}, with a characteristic minimum in the transition region between the material strength regime and the gravity regime (at $\sim 10$...$100\m$). At the very right lie the planetesimals that have not reached their collisional lifetime yet, which therefore retain their initial (primordial) distribution.
Ripples seen above the blowout limit and above the transition to gravity-dominated regimes at $10^7-10^8\mum$, especially for the \fomaref run, are a known phenomenon \citep[e.g.,][]{campo-et-al-1994b,wyatt-et-al-2011} and will not be discussed further.

As a side remark, the above analysis also confirms that the same dust mass at a given age can result from a population of planetesimals of a lower initial mass if $\QD(s)$ is steeper, as expected~--- potentially mitigating the ``debris disk mass problem'' \citep{krivov-et-al-2018,krivov-wyatt-2021}. For instance, our \zodisteep run with just $0.32M_\oplus$ in planetesimals sustains the same amount of dust at $100\Myr$ as the \reference run disk with the total mass of $100 M_\oplus$ (cf. blue and red solid lines in Fig.~\ref{fig:Mdust}).  It is highly questionable, however, whether a non-traditional $\QD$ model such as that of the \zodisteep run is a solution to this problem. The more so, since it is not clear yet whether simulation results for alternative models are compatible with other debris disk observables, such as radial profiles of brightness at different wavelengths. This will be considered in the subsequent sections.

\subsection{Spatial distribution}
\label{spatialDistribution}

\begin{figure}[ht!]
    \centering
    \includegraphics[width=1.0\linewidth]{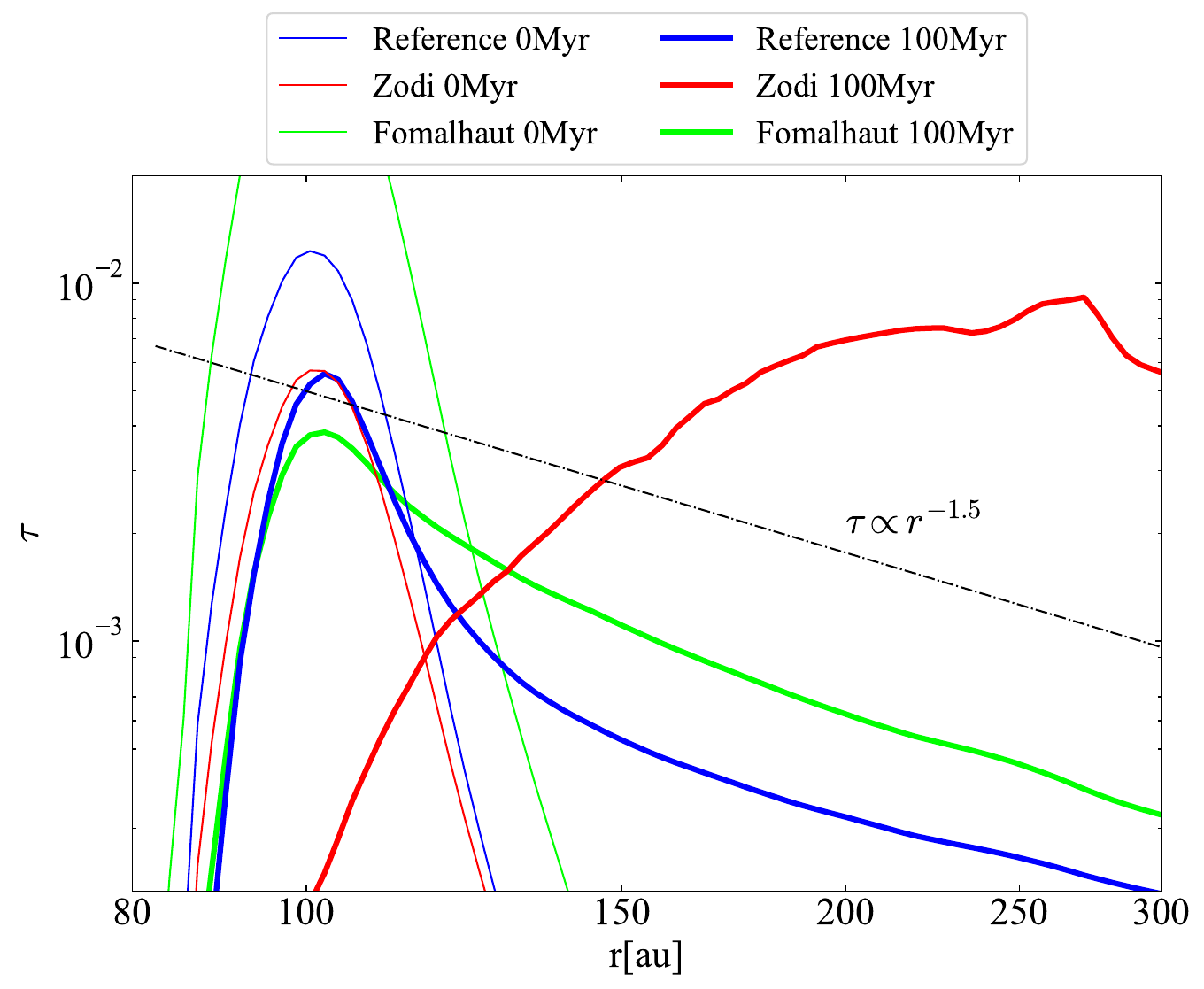}
    \caption{
    Radial profiles of normal geometrical optical depth in the three runs. An idealized profile (eq.~\ref{tau1.5}) and is shown for comparison. Both axes are logarithmic.
    }
    \label{fig:tau}
\end{figure}

We now look at the optical depth profiles of the simulated debris disks (Fig.~\ref{fig:tau}).
For an idealized disk, the normal geometrical optical depth $\tau$ should decay at greater distances to the star as \citep{strubbe-chiang-2006,krivov-2010}
\begin{equation}
    \tau(r)\propto r^{-1.5}.
    \label{tau1.5}
\end{equation}
While the optical depth profiles in the \reference and \fomaref runs do converge to this idealized one, the profile in the \zodisteep run differs dramatically. 
In this case, the optical depth increases farther out from the star, which is totally unexpected and needs to be understood. In addition, the profile is not quite smooth, exhibiting, for instance, some dips at $\approx160\au$ and $\approx240\au$. This behavior arises from the discrete phase space grid of the numerical simulation.

\begin{figure*}[htb!]
    \centering
    \includegraphics[width=\linewidth]{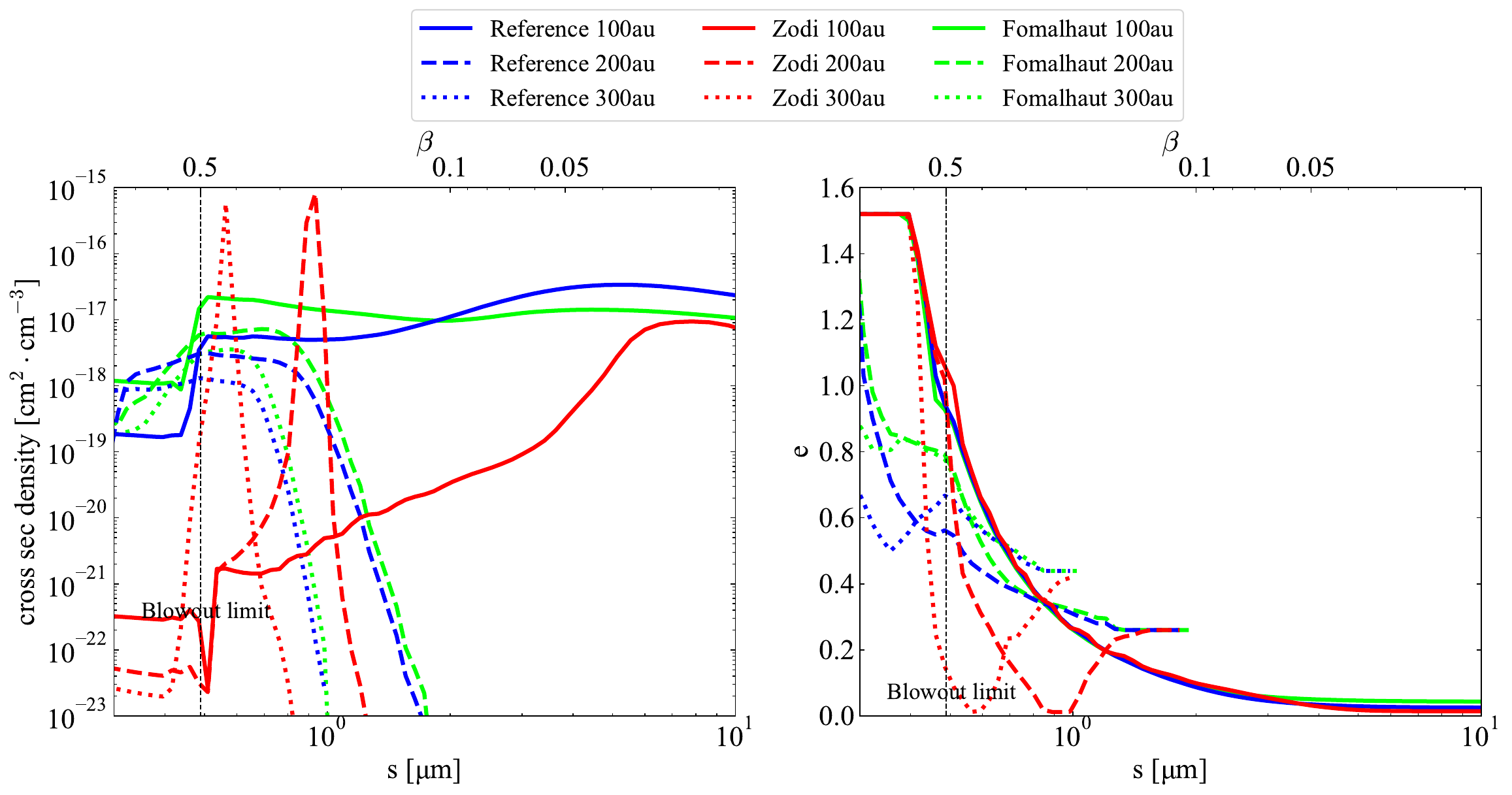}
    \caption{
    Size distribution of dust grains (left) and average eccentricity of different-sized dust grains (right) in the \reference, \fomaref, and \zodisteep runs at three selected stellocentric distances: $100\au$ (solid lines), $150\au$ (dashed), and $200\au$ (dotted). Note that only dust grains smaller than $10\mum$ are shown. In both panels, the blowout grain size is marked with a vertical black line.
    }
    \label{fig:CrossSec200au_Ecc}
\end{figure*}

To find a physical mechanism behind the increasing optical depth profile
in the \zodisteep run, we first looked at the size distribution of dust at different locations, both in the parent ring and exterior to it. The left panel of Figure~\ref{fig:CrossSec200au_Ecc} shows the size distribution for $s<10\mum$ in all three runs at $100\au$, $200\au$, and $300\au$. 
For the \reference and \fomaref runs, the $100\au$ distribution extends across the entire size range, the $200\au$ distribution has a significant cross-section density between somewhat below the blowout size and $\approx 2\mum$, and the $300\au$ distribution between the minimum size and $\approx 1\mum$.
However, the \zodisteep run is very different. The $100\au$ distribution has a broad peak at
$s \approx 5\mum$, and the $200\au$ and $300\au$ distributions feature very sharp peaks at $0.9\mum$ and $0.6\mum$, respectively. This suggests that in the \zodisteep run, each radial location in the outer disk is preferentially populated by grains of specific sizes. While size sorting is a known phenomenon in debris disks \citep{strubbe-chiang-2006,thebault-wu-2008,thebault-et-al-2014}, the effect is much more pronounced in the \zodisteep run than in the \reference and \fomaref runs.

Next, we checked the average orbital eccentricity of different-sized dust grains in the same runs and at the same three distances as before. The results are shown in the right panel of Fig.~\ref{fig:CrossSec200au_Ecc}. We see that in the \zodisteep run, the size distributions in the outer disk (at $200\au$ and $300\au$) peak at the grain sizes at which the average eccentricity reaches its minimum. This implies that specific regions in the outer disk exist where particles of certain sizes accumulate on nearly circular orbits.

\begin{figure}[htb!]
    \centering
    \includegraphics[width=0.8\linewidth]{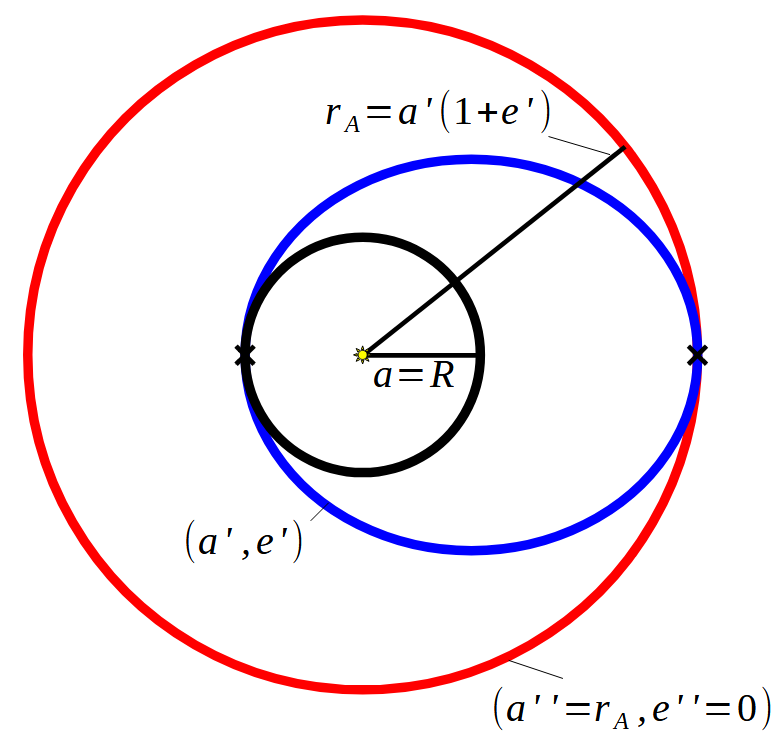}
    \caption{
    A schematic of the proposed circularization mechanism. Black: the parent belt, blue: a halo grain, red: a secondary debris particle in a circular orbit.
    }
    \label{fig:OrbitSchematic}
\end{figure}

Trying to explain why in the \zodisteep run small grains concentrate in nearly circular orbits on the outer disk, we made additional test runs with different combinations of collisional outcomes described in Section~\ref{Sec:outcomes}. The results are described in Appendix~\ref{sec: app rebounding}. We found that excluding rebounding collisions (eqs.~\ref{eq:reb1}--\ref{eq:reb3}) changes the evolution dramatically (see Appendix~\ref{sec: app rebounding}). This is illustrated by
Fig.~\ref{fig:tauNoRebounding} (radial profiles of optical depth), Fig.~\ref{fig:CrossSecNoRebound} (size distribution), and Fig.~\ref{fig:EccNoRebounding} (average eccentricity of grains of different sizes).
Excluding cratering collisions in addition to rebounding collisions brought about results that were almost indistinguishable from those with just the rebounding collisions removed. No runs were performed without sticking collisions, as they occur too rarely to make a difference.

Accordingly, we now focus on how rebounding collisions in the \zodisteep run circularize orbits of dust grains of specific sizes. Analysis of the simulation bins uncovered the following picture (Fig.~\ref{fig:OrbitSchematic}):
\begin{itemize}
\item
Collisions in the parent ring at $R=100\au$ generate dust. Grains with a certain size $s$ and $\beta=\beta(s)$ are sent by radiation pressure in eccentric orbits with semimajor axes and eccentricities \citep{burns-et-al-1979}
\be
a^\prime = R (1-\beta)/(1-2\beta)
\quad \mathrm{and} \quad 
e^\prime = \beta/(1-\beta),
\ee
respectively. The pericenter of these orbits lies in the parent ring at $r_\Pi = R$, and apocenter is at $r_\mathrm{A}= R/(1-2\beta)$. These grains form well-known halos exterior to the parent belt.
\item
Around the apocenter, rebounding collisions between those grains and highly eccentric particles with pericenters in the main belt produce yet smaller dust debris. Again, radiation pressure sends these to new orbits with semi-major axes $a^{\prime\prime}$ and eccentricities $e^{\prime\prime}$, depending on their $\beta$-ratio. Since the release point of those secondary debris is at the apocenter, there exist some sizes $s^\prime$ (and $\beta^\prime$) such that the new debris gets on circular orbits with $a^{\prime\prime}=r_\mathrm{A}$ and $e^{\prime\prime}=0$. The enhancement of density (or optical depth) at distance $r$ caused by grains on circular orbits is a simple geometrical effect \citep[see, e.g.][]{haug-1958}.
\end{itemize}
Consider now a certain distance $r$ in the outer disk (e.g., one of the distances at which an enhanced optical depth is seen in Fig.~\ref{fig:tau}). It is easy to derive both the $\beta$-ratio of the ``primary'' ejecta from the parent ring with apocenters at distance $r$,
\be
\beta = \frac{1}{2}\left( 1 - \frac{R}{r}\right) ,
\label{betaclassic}
\ee
and that of the ``secondary'' debris arising from rebounding collisions and placed into circular orbits or radius $r$,
\be
\beta^\prime = \left( 1 - \frac{R}{r}\right) = 2\beta.
\label{betaprime}
\ee

\begin{figure}[htb!]
    \centering
    \includegraphics[width=1.0\linewidth]{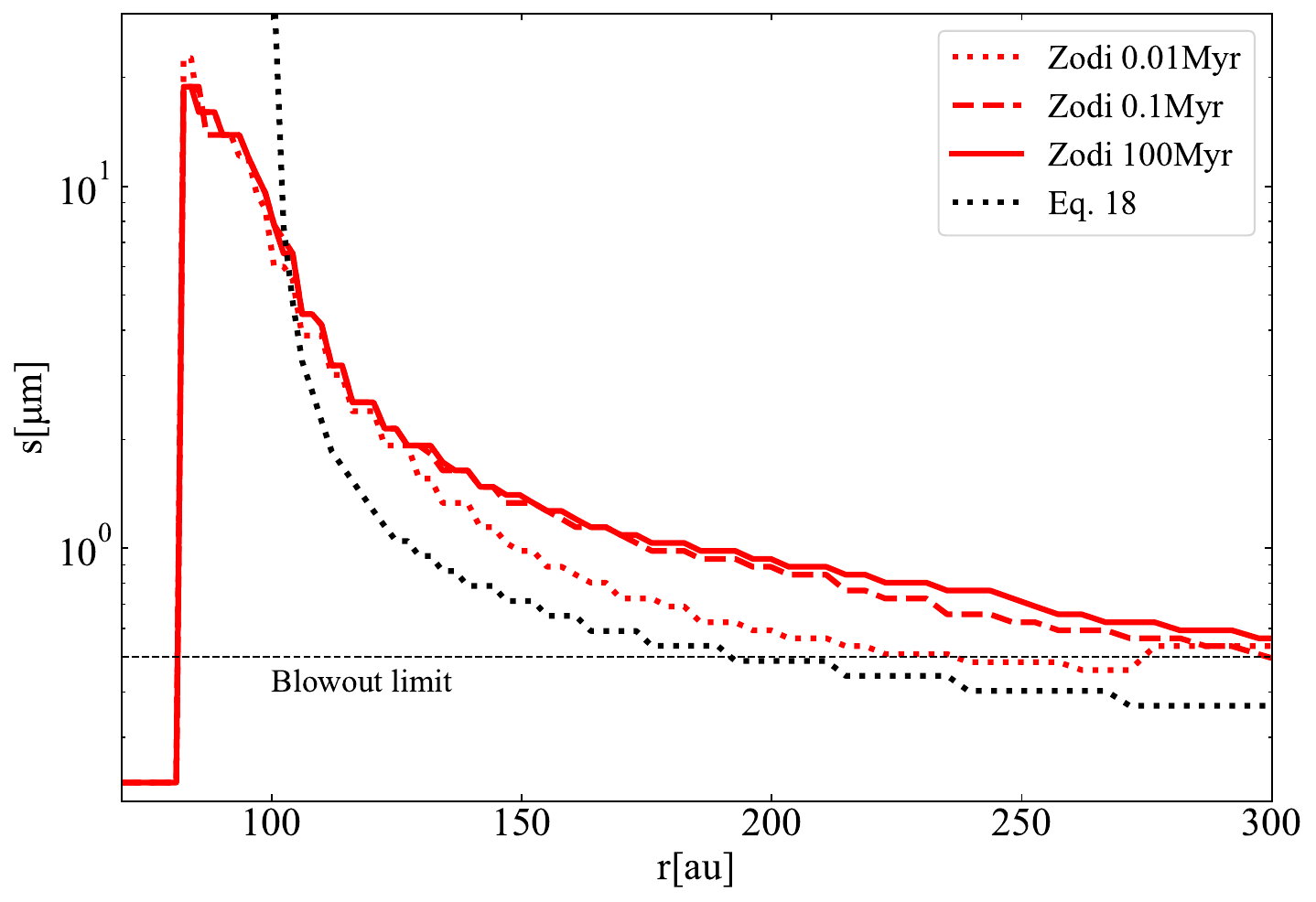}
    \caption{
    Peak positions of cross-section density distributions of the \zodisteep run as functions of distance from the star at three different system ages.
    Overplotted in black is the theoretical curve, eq.~(\ref{betaprime}).
    }
    \label{fig:SizeDependentSigma}
\end{figure}

We now make simple numerical estimates.
At $r=200\au$, we obtain $\beta = 0.25$ and $\beta^\prime = 0.50$, which translate to sizes $s=0.9\mum$ and $s^\prime = 0.5\mum$. At $r=300\au$, we get $\beta = 0.33$ and $\beta^\prime = 0.67$, which correspond to sizes $s=0.7\mum$ and $s^\prime = 0.4\mum$.
The sizes $s^\prime$ deviate somewhat from the positions of the eccentricity minima seen in Fig.~\ref{fig:CrossSec200au_Ecc}, which are $0.9\mum$ and $0.6\mum$, respectively. Further investigation revealed that while the size-dependent radial location of the circularized dust does not perfectly match the predictions of eq.~(\ref{betaprime}) at $100\Myr$, it does at beginning of the collisional evolution. To demonstrate this, in Fig.~\ref{fig:SizeDependentSigma} the positions of the peaks of the size distributions of the \zodisteep run versus stellocentric distance are plotted at system ages of $0.01\Myr$, $0.1\Myr$, and $100\Myr$. The expected grain sizes from eq.~(\ref{betaprime}) are overplotted for comparison. It is evident that, at $0.01\Myr$, the \zodisteep run matches eq.~(\ref{betaprime}) reasonably well, but starts to deviate from it quickly, with even the $0.1\Myr$ distribution being very different from it. This suggests that the circular dust orbits in the \zodisteep run do appear at the positions predicted by eq.~(\ref{betaprime}) and later ``migrate'' outwards.
Analyzing the simulation bins, we conclude that expansion of the circular dust orbits is caused by a cascade of rebounding collisions which gradually erode the grains, reducing their sizes, and shift them to greater radii.
Further details and plots illustrating the circularization of dust orbits in the \zodisteep run are given in Appendix~\ref{sec: app circular rings}.

The question arises why we do not see such circular orbits in the \reference and \fomaref runs. In all three runs, dust particles born in the main belt are placed in orbits with high apocenters by radiation pressure. However, only in the \zodisteep run, rebounding collisions dominate, which reduce particle sizes more gently than disruptive collisions would do, producing grains with ``right'' sizes (eq.~\ref{betaprime}) to get in circular orbits. In the \reference and \fomaref runs, however, these particles get destroyed via disruptive collisions with high-eccentricity grains stemming from the main belt. The fragments of these collisions are too small to be placed by radiation pressure in circular orbits. Instead, they typically get in hyperbolic orbits and are blown out of the system. If any particles on circular orbits are produced, they are quickly destroyed by subsequent collisions.

In summary, we found that a disk of hard dust, represented by our \zodisteep run, contains a large amount of small dust particles in circular orbits at large stellocentric distances. These are produced by rebounding collisions in the outer disk and, due to their high $\QD$ value, survive subsequent collisions with other particles. These small dust grains in circular orbits are so numerous that the cross-section density in the outer disk gets larger than in the main belt (see Fig.~\ref{fig:CrossSec200au_Ecc} left).
It is this phenomenon that explains the increase of the optical depth with distance seen in Fig.~\ref{fig:tau}. 

\subsection{Spectral energy distributions}
\begin{figure}[ht!]
    \centering
    \includegraphics[width=1.0\linewidth]{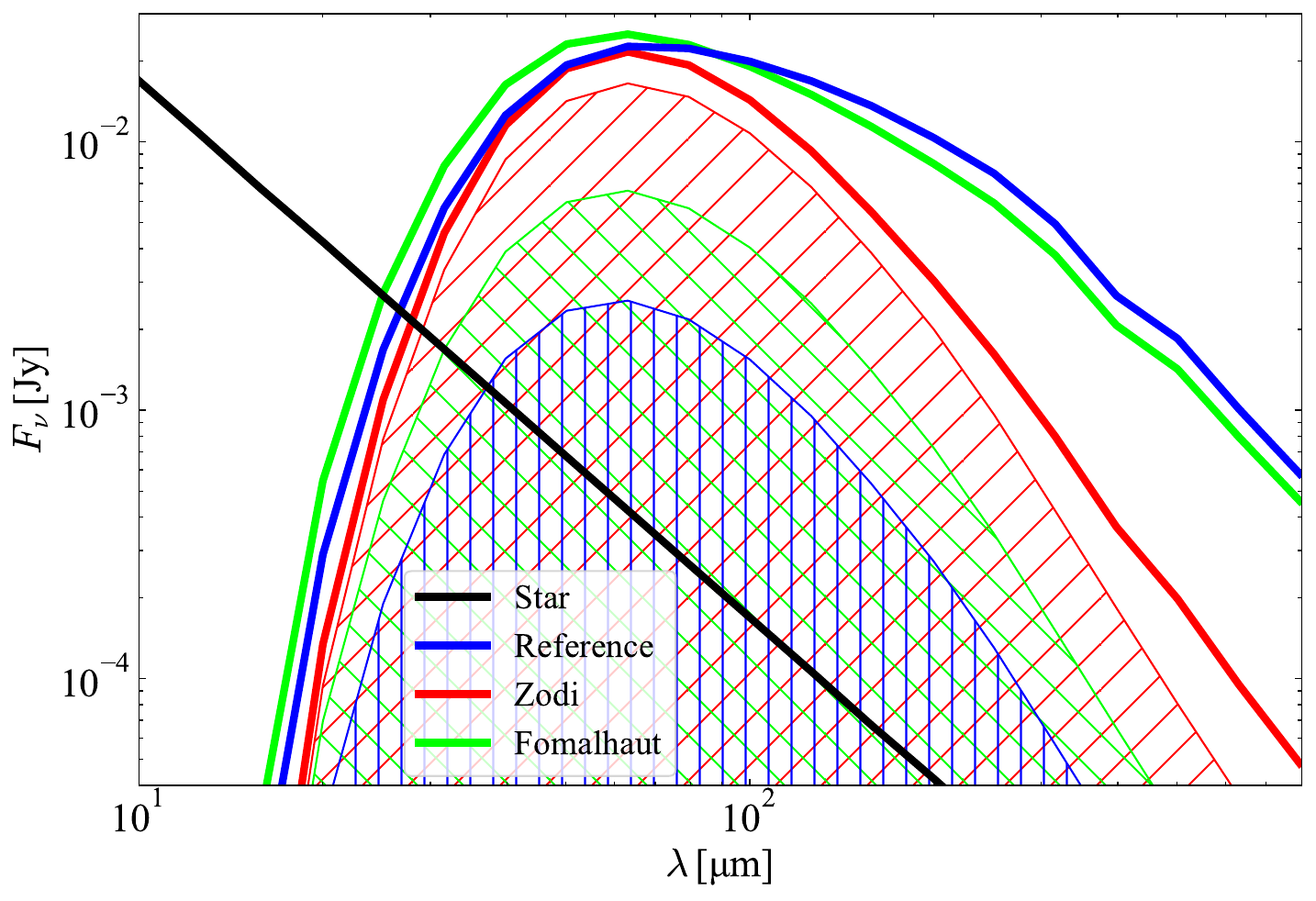}
    \caption{
    Thermal emission SEDs in the three runs analyzed (colored lines). The hatched regions indicate the portion of the SEDs generated by the part of the disk outside of $150\au$. The black line is the stellar photosphere.
    }
    \label{fig:SED}
\end{figure}

Our intention in this study is to find out whether one can distinguish between different $\QD$ models from debris disk observations. The simplest ``observable'' for a debris disk is its spectral energy distribution (SED). However, the SEDs derived for our three runs turn out to be pretty similar (Fig.~\ref{fig:SED}). This might not be intuitive as the size distributions differ markedly (Fig.~\ref{fig:SizeDist}). In particular, \zodisteep run's distribution peaks at a larger particle size than the \reference run's. Since larger particles at the same distance are colder, we should expect its SED to peak at a longer wavelength. However, Fig.~\ref{fig:SizeDist} only shows the size distribution at $100\au$, whereas the \zodisteep disk contains many more small particles at larger stellocentric distances. The net effect is that the SEDs are roughly the same. Indeed, most of the \zodisteep run's SED originates from high stellocentric distances and not from the main belt. This can be seen in Fig.~\ref{fig:SED}, where the hatched area represents the part of the SED which is generated by the outer disk at $r>150\au$. While its contribution can be neglected for the \reference and \fomaref runs, it makes up most of the SED for the \zodisteep run. 

To compare the SEDs quantitatively, a useful quantity
is $\Gamma$, the ratio of the dust belt radius $R$ and its blackbody radius $R_\text{bb}$ \citep{booth-et-al-2013,pawellek-et-al-2014,pawellek-krivov-2015}:
\begin{equation}
    \Gamma = R/R_\text{bb},
\end{equation}
where $R=100\au$ in our simulations, and the blackbody radius is given by \citep{backman-paresce-1993}
\begin{equation}
    R_\text{bb} = (278\K/T_\text{dust})^2\cdot(L_*/L_\odot)^2,
\end{equation}
with $L_*$ being the stellar luminosity and $T_\text{dust}$ the dust temperature,
\begin{equation}
    T_\text{dust} = \frac{5100\K\mum}{\lambda_\text{max}},
\end{equation}
which is computed from the peak wavelength $\lambda_\text{max}$ of the SED (assuming that thermal emission flux is given in Jy, i.e. is measured per unit frequency).

The thermal emission of all three dust belts peaks at $\lambda_\text{peak}=63\mum$ which, using the above formulae, corresponds to a black body equilibrium temperature of $T=81\K$ and a blackbody radius of $r_\text{bb}=12\au$. The resulting $\Gamma=8.3$ is therefore the same for all three runs, not allowing one to distinguish between different $\QD$ prescriptions based on a disk's SED.

\subsection{Radial brightness profiles}
\label{subsec: brightness}

\begin{figure}[h!]
    \centering
    \includegraphics[width=1.0\linewidth]{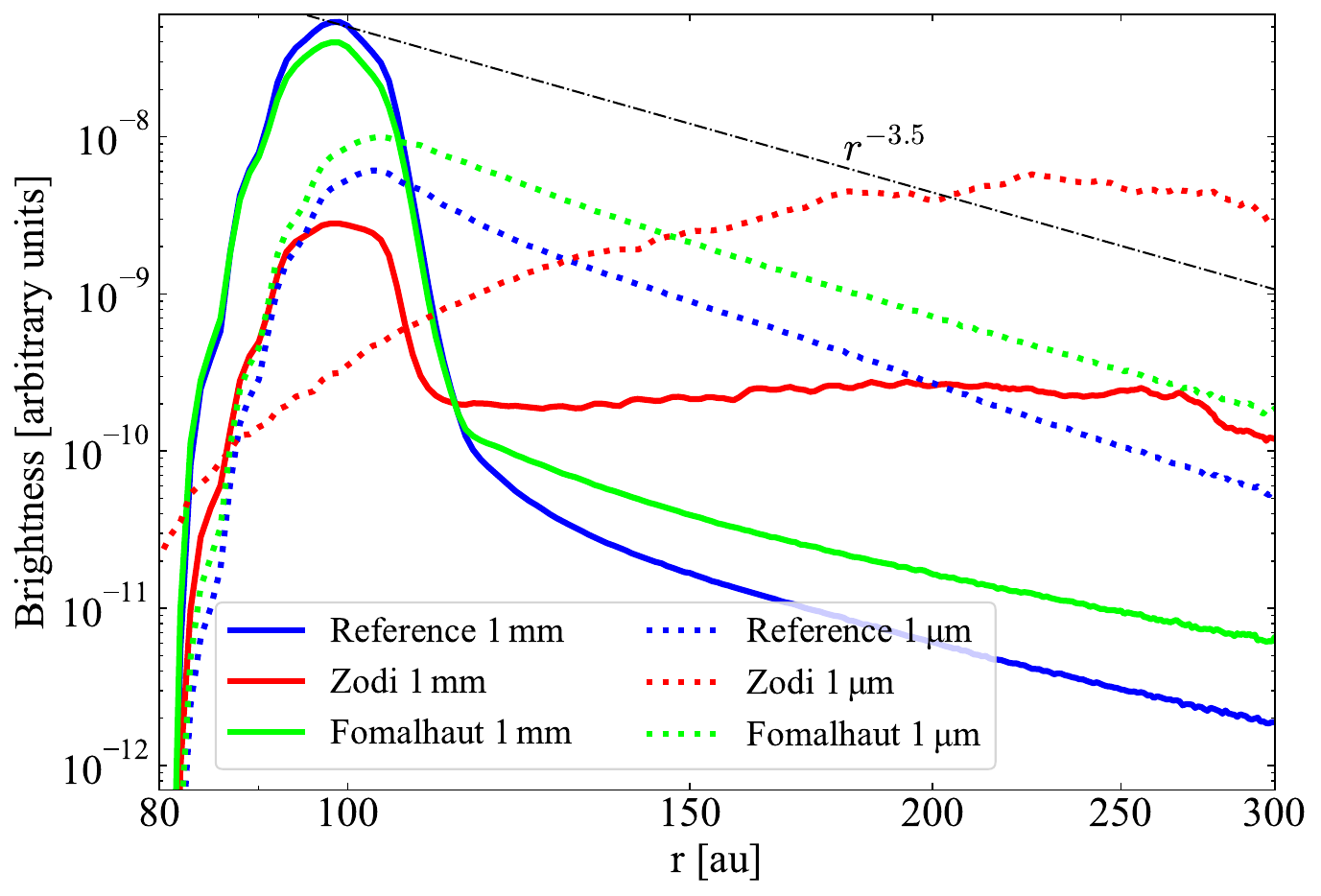}    \caption{
    Profiles of disk brightness (per pixel) in the three runs
    after $100\Myr$ of evolution, at wavelengths of  $1\mum$ (dashed) and $1\mm$ (solid lines), using a resolution of $1\aupx$.
    }
    \label{fig:BrightnessFINE}
\end{figure}

While the differences in the optical depth profiles (Fig.~\ref{fig:tau}) are interesting, more promising for our purposes are brightness profiles at different wavelengths. We computed them as explained in Section~\ref{subsec: longterm}.
These are shown in Fig.~\ref{fig:BrightnessFINE} for the three runs after 100\,Myr, both in scattered light at $1\mum$ and in thermal emission at $1\mm$. 

For the sake of clarity,
we only consider here the simplest case of a debris disk with a single, narrow parent belt (see the setup of our simulations, Section~\ref{subsec: setup}). A significant fraction of the observed disks do belong to this category \citep{han-et-al-2026}. We note, however, that disks with a more complex radial structure exist as well. Some appear broad at mm wavelengths \citep[e.g.][]{matra-et-al-2025,marino-et-al-2026}, while some others contain multiple parent rings, often both narrow and broad \citep{han-et-al-2026}. There are even atypical systems such as HD~131835 \citep{jankovic-et-al-2026} where the mm and scattered-light profiles peak at completely different distances from the star. Analysis of such systems is out of the scope of this paper and would require a separate study.

In thermal emission, brightness profiles of debris disks with a single, narrow parent belt are observed to show a pronounced peak around the parent belt, followed by a falloff farther out from the star that often drops quickly to below the noise level \citep[e.g.,][]{marino-2021,han-et-al-2026}. In scattered light, a peak nearly at the same position\footnote{In some debris disks with CO gas detections, the scattered light intensity peaks somewhat farther out from the star than the thermal emission does, which is attributed to outward drift of small $\mum$-sized grains by gas drag forces \citep[see][and Olofsson et al., A\&A, submitted]{milli-et-al-2026}.} as with thermal radiation is seen, followed by a gentle decrease in brightness, often~-- although not always~-- approximately
as $r^{-3.5}$ \citep{thebault-wu-2008,thebault-et-al-2023,engler-et-al-2025,milli-et-al-2026}.
Figure~\ref{fig:BrightnessFINE} readily demonstrates that both \reference and \fomaref profiles are fully compatible with these expected brightness profiles.
It would be difficult to distinguish between the two models, \reference and \fomaref, given numerous and diverse uncertainties, ranging from unknown distributions of dust-producing planetesimals to challenges of finding the phase function that reproduces scattered light data adequately.

While the \reference and \fomaref runs yield reasonable results fully compatible with typical debris disk data, the \zodisteep profiles are completely different. The only similarity of the thermal emission profile with actual data is that it peaks at the parent ring. Exterior to it, the brightness profile is pretty flat. This might seem unexpected given that $1\mm$ flux is typically generated mostly by dust roughly of the size of $1\mm$, and that there is no such dust at distances of $200\au$ or greater (see Fig.~\ref{fig:CrossSec200au_Ecc}). In this case, however, the $1\mm$ flux is generated by the very small $\approx1\mum$ dust which is abundant at those distances. While its absorption and emission efficiency is very low, $Q_\text{abs} \sim10^{-4}$ using Fig.~6 from \citet{krivov-et-al-2008}, the amount of small dust is so high that the $\mm$ flux in the outer disk stays nearly constant. The scattered-light profile of the \zodisteep run is pathological as well, and clearly at odds with the actually observed profiles. No traces of the parent ring are visible at all, and the brightness increases outward. 

\section{Discussion}
\label{Sec:Discussion}

\subsection{When is dust too hard?}
In Sections~\ref{spatialDistribution} and \ref{subsec: brightness}, we found that lowering the $\QD$ value of dust particles below the value of the \reference run does not significantly alter the radial profile of a debris disk, while increasing $\QD$ leads to peculiar profiles that are clearly incompatible with observations. This raises a question: when is dust too hard?  

From Section~\ref{spatialDistribution} and the analysis in Appendix~\ref{sec: app circular rings}, we know that peculiar radial profiles arise if small particles in the outer disk are placed in circular orbits. This was the case in the \zodisteep run.
And conversely, orbits of small dust grains in the outer disk do not get circularized, if their source~-- halo particles with pericenter in the main belt and apocenters far away from the star~-- get destroyed by near-blowout particles created in the main belt. In this case, exemplified by the \reference run, the radial profiles are ``normal'' and consistent with the observed ones.

This allows us to derive a simple criterion for the absence or presence of dust grains in circular orbits in the outer disk.
Analyzing the results of the \reference run, we found that only 10-20\% of collisions between the near-blowout particles and the halo ones occur in the main ring, while half of them happen far away from it, closer to the apocenters of the halo grains.
To make a rough estimate, we assume that the collisions take place at the apocenters. Further, we assume that small projectiles, which are created at $R$ and are to destroy a particle of size $s\sbs{t}$ in an eccentric orbit with apocenter at $r$, the splinter fragments of which would get in a circular orbit of radius $r$, have $\beta=1/2$ and so exactly the blowout size $s_\text{blow}$. They move away from the star in a parabolic orbit, so that their velocity at the distance $r$ is
\begin{equation}
 v = v_\text{K}(R) \sqrt{R/r} = v_\text{K}(r) ,
\label{vK}
\end{equation}
where $v\sbs{K}(r)$ is the circular Keplerian speed at a distance $r$.
Finally, we assume that the relative velocity between the target particles in eccentric orbits and projectiles in parabolic ones is approximately equal to the speed of the latter, $v$.

The maximum critical fragmentation energy $\QD$ for which the collision will be disruptive is given by \citep[see eq.~(5.3) in][]{krivov-et-al-2005}
\begin{equation}
    \QD(m\sbs{t}) = \frac{1}{2}v^2\frac{m\sbs{blow}}{m\sbs{t}} 
\label{QD crit general}
\end{equation}
or
\begin{equation}
    \QD(s\sbs{t}) =
    \frac{1}{2}v^2
    \left( \frac{s\sbs{blow}}{s\sbs{t}} \right)^3,
\label{QD crit general s}
\end{equation}
where $m\sbs{blow}$ and $m\sbs{t}$ are the masses of the blowout-sized bullets in parabolic orbits and the target grains in eccentric orbits, respectively. Equations~(\ref{QD crit general}) and (\ref{QD crit general s}) are valid for $m\sbs{blow} \ll m\sbs{t}$.

We now express $s\sbs{t}$ through the distance $r$. Since grains of size $s\sbs{t}$ are source of grains in circular orbits at the distance $r$, their $\beta$ is given by eq.~(\ref{betaclassic}). Assuming that radiation pressure efficiency for particles with $s\sbs{t}>s\sbs{blow}$ is unity, $\beta \propto 1/s$, and we write
\begin{equation}
        s\sbs{t} = \frac{1}{2\beta} s\sbs{blow} ,
        \label{s-target}
\end{equation}
so that
\begin{equation}
    s\sbs{t}=\frac{s\sbs{blow}}{1-R/r} .
\label{st}
\end{equation}

With the above assumptions and using eqs.~(\ref{vK}) and~(\ref{st}), eq.~(\ref{QD crit general s}) takes the form:
\begin{equation}
    \QD(s\sbs{t}) = \frac{1}{2}\,v\sbs{K}^2(r)\,
    \left( 1 - \frac{R}{r} \right)^3 .
\label{QD crit}
\end{equation}

For any distance $r>R$ in the outer disk, eq.~(\ref{st}) gives the size $s\sbs{t}$, and eq.~(\ref{QD crit}) gives the maximum $\QD$ that dust grains of that size are allowed to have in order not to send splinters to circular orbits and not to form abnormal radial profiles.

Since the above derivation assumes $m\sbs{blow} \ll m\sbs{t}$, the model is no longer valid for $r \gg R$. In that case, a general eq.~(5.2) from \citet{krivov-et-al-2005} instead of eq.~(5.3) shall be used. For large $r$, eq.~(\ref{st}) becomes \begin{equation}
    s\sbs{t} \approx s\sbs{blow},
\label{st far}
\end{equation}
and eq.~(\ref{QD crit}) replaces by
\begin{equation}
      \QD(s\sbs{t}) \approx \frac{1}{8}\,v\sbs{K}^2(r) .
\label{QD crit far}  
\end{equation}

Let us make numerical estimates with eqs.~(\ref{st})--(\ref{QD crit}), which are valid at distances within a few $R$.
As before, we take $R=100\au$, $v(R)=3\kms$, and $s\sbs{blow}=0.5\mum$.
At $r=2R=200\au$, we get $s\sbs{t}=1\mum$ (close to the exact value $0.9\mum$ which comes from the actual $\beta(s)$ dependence that assumes a realistic radiation pressure efficiency) and $\QD \approx 6\times10^9\ergg$.
This is larger than $\QD$ at $0.9\mum$ of the \reference run, $7\times10^8\ergg$, and smaller than in the \zodisteep run, $2\times10^{10}\ergg$.
Similarly, at $r=3R=300\au$, the result is $s\sbs{t}= 0.75\mum$ and $\QD \approx 7\times10^9\ergg$. Again, this is larger than $\QD(0.75\mum)$ in the \reference run and smaller than in the \zodisteep one.
At large distances, such as $r=10R=1000\au$, we use eqs.~(\ref{st far})--(\ref{QD crit far}) to get $s\sbs{t} \approx 0.5\mum$ and $\QD \approx 1\times10^9\ergg$.
These estimates are fully consistent with that fact that circular orbits do appear in the \zodisteep run, but get destroyed in the \reference run.

The above analysis gives the strength threshold above which small grains in the outer disk are sent into circular orbits. Strictly speaking, this alone may not be sufficient for peculiar brightness profiles to arise. It can happen that circular orbits form, but get swiftly destroyed~-- by the same near-blowout particles produced in the main ring. To check when this happens, we start again with the relative velocities between the target grains in circular orbits and near-blowout grains impacting them. Within a factor of two, these relative velocities are still given by eq.~(\ref{vK}). Further, particles in circular orbits typically have sizes close to the blowout one (see Section~\ref{spatialDistribution}). Therefore, we now have the case $s\sbs{t} \approx s\sbs{blow}$, which is described by eq.~(\ref{QD crit far}). We conclude that the criteria for creation and survival of dust grains in circular orbits are almost the same. For ball-park estimates, one can simply use eqs.~(\ref{st far})--(\ref{QD crit far}).

\subsection{Caveats}

In our simulations, several physical effects of potential importance
were omitted. One of the them is the Poynting-Robertson (PR) effect. The PR effect can often, although not always, be ignored, especially in bright disks that have high optical depth \citep{wyatt-2005,kennedy-piette-2015,rigley-wyatt-2020,sommer-et-al-2025}. This is because the collisional lifetimes of dust particles are typically much shorter than their PR lifetimes. However, in this work's \zodisteep run, collisions are not necessarily catastrophic anymore and, at least in the outer region of the disk, just slightly erode the particles and stochastically change their orbits. As a result, grain lifetimes in disks composed of hard dust may become much longer. Therefore, including PR effect in simulations seems reasonable. Accordingly, we made an additional test run with the same setup as in the \zodisteep one, but with the PR drag switched on.
The differences to the normal \zodisteep run without PR drag turned out to be only marginal.

Nevertheless, including the PR effect is absolutely necessary for debris disks with low fractional luminosities such as the Solar system's zodiacal dust cloud. Therefore, we performed one more \zodisteep run with PR drag included and an initial disk mass set to an arbitrary low value ($10^{-4}\,M_\oplus$).
This run did not reveal the pathological optical depth profile seen in the \zodisteep run. Instead, the profile turned out to be perfectly consistent with the one typical of transport-dominated disks: $\tau \propto r^{-2.5}$ exterior to the parent ring and $\tau \propto r^0$ interior to it \citep{strubbe-chiang-2006}.
We conclude that the constraints on $\QD$ obtained in this paper do not necessarily apply to disks with low optical depths. This might also be part of the explanation why \citet{rigley-wyatt-2022} found hard dust to fit well the zodiacal cloud data (see also discussion in Section~\ref{rebound exist}). Including the PR effect would be required to place constraints on the critical fragmentation energy of dust in such disks. Unfortunately, handling transport processes like PR drag is more expensive computationally than that of pure collisional evolution. Also, the mass scaling (eq.~\ref{shiftLaw}) is not valid in the presence of transport, which would further hamper the analysis. 

Another effect not included in our modeling is collisional damping \citep[][]{pan-schlichting-2012}.
This might become important in realistic disks with hard dust \citep{jankovic-et-al-2024}. If so, damping would also affect the vertical velocities, flattening the disks. Such effects cannot be modeled with our collisional code ACE, as it assumes the vertical velocities to be constant and does not resolve the inclination.

\subsection{Do rebound-dominated disks exist?}

\label{rebound exist}
In collisional modeling of debris disks, it is typically assumed that disks are dominated by disruptive collisions \citep[see, e.g.,][among others]{dohnanyi-1969, wyatt-et-al-1999, wyatt-dent-2002}. \citet{thebault-et-al-2003} and \citet{kobayashi-tanaka-2010} pointed out that cratering collisions can be even more important in many cases. In this work,
we found that rebounding collisions (those in which both impactors bounce off each other and survive, only experiencing some erosion) may play the leading role in some disks.

Reboundung (or bouncing) collisions have been in the focus of interest for protoplanetary disks, and were intensively studied in laboratory experiments and theoretical studies of the early stages of planet formation \citep[e.g.,][]{guettler-et-al-2010}. For instance, the ``bouncing barrier'', which describes the inability of solids to grow further beyond some sizes or velocities due to the repeated bouncing events, is a well-known challenge of planet formation theory \citep[e.g.][among others]{zsom-et-al-2010,windmark-et-al-2012,bukhari-et-al-2017}. However, to our knowledge, this type of collisional outcome has never been considered in the context of protoplanetary disk descendants, debris disks.
A natural question is: do such rebound-dominated debris disks exist in reality?
In other words, are there any conditions under which $\QD$ of dust grains would lie above the critical value (eq.~\ref{QD crit} and eq.~\ref{QD crit far})?
These equations suggest that, somewhat counterintuitively, the critical value of $\QD$  does not depend on the stirring level of the disk material (i.e., the eccentricity and inclination dispersion of solids or, equivalently, their relative velocities). Instead, it only depends on the local circular Keplerian velocity at the debris location. This means that the smallest critical values are expected where $v\sbs{K}$ is the lowest, i.e. in the outermost parts of large disks~-- or around very low-mass stars. However, that dependence is not strong: the critical $\QD$ is proportional to the square root of the stellar mass and inversely proportional to the square root of distance. Therefore, the expected critical value could only be lowered by a factor of several~-- even on the outskirts of large disks of low-mass stars.
Another logical possibility would be to find debris disk systems in which the dust itself is mechanically much harder than in classical exo-Kuiper belts. One might think of some material compositions; also, this would certainly require the absence of any micro-porosity. However, it is not clear whether, and in which systems, this could be the case.

Remember that our \zodisteep model was motivated by the results of \citet{rigley-wyatt-2022} who found a steep $\QD$~-- implying hard micrometer-sized dust~-- to fit best the zodiacal dust cloud data in their model. The reasons for such a $\QD$ are explained in their Section 4.2. Considering dust produced by cometary sources and its subsequent evolution under PR drag and collisions, they found that in a canonical model (similar to our \reference one) the radial slope of the optical depth in the dust cloud ($-0.45$) was steeper than that inferred from the data \citep[$-0.34$,][]{kelsall-et-al-1998}. A steeper $\QD(s)$ dependence allowed them to skew the size distribution towards larger sizes and to flatten the radial slope, achieving a good fit to the data. We note, however, that this was done based on the model that only included disruptive collisions. Any collision with energy above the disruption threshold was assumed to eliminate the colliders from the system. And conversely, any collision with insufficient impact energy left the colliding grains intact. It would be interesting to see how the results of their modeling would be affected if additional collisional outcomes, especially rebounding collisions, were implemented and the fragments of such collisions were taken into account. On the other hand, the \citet{rigley-wyatt-2022} model did include the PR effect, which our simulations did not. Hence it would be desirable to re-do a similar analysis both with realistic collisional outcomes and PR drag switched on.

Although it is questionable that rebound-dominated debris disks exist, and actually because of that, this regime of collisional evolution is as yet unexplored and lacks theoretical groundwork, similar to the one laid over the last decades for the disruption- and cratering-dominated regimes. Analysis of circular orbits of small particles with certain sizes that form at large stellocentric distances performed in Section~\ref{spatialDistribution} is the first step.
Moreover, future research in this direction may provide further insights into disruption- or cratering-dominated disks, since rebounding collisions are very common in those as well.

\subsection{Can this analysis help solve the ``mass problem''?}

Our attempts to constrain the mechanical strength of dust might have some implications to the ``mass problem of debris disks'' described in \citet{krivov-et-al-2018,krivov-wyatt-2021}. Using standard prescriptions for $\QD$ similar to this work's \reference run and a maximum planetesimal size of $\approx100\km$, one can calculate the total mass of a debris disk just using its dust mass inferred from observations. For many debris disks, this estimate results in disk masses greater than $1000\,M_\oplus$, which exceeds the mass of solids available for planetesimal formation at the preceding, protoplanetary stage of systems' evolution.
One possible solution to this problem is to reduce the maximum size of planetesimals, as most of the disk mass resides in the largest planetesimals \citep{krivov-wyatt-2021}. Another one would be a steeper $\QD(s)$ dependence than usually assumed. This is because a steeper $\QD$ prescription
would lead to a steeper size distribution \citep{o'brien-greenberg-2003},
thus requiring fewer large planetesimals to sustain the same amount of the observed dust. Indeed, we saw that in our \zodisteep run with its steeper $\QD$ in the strength regime, a total disk mass of $0.3M_\oplus$ in bodies up to $200\km$ in size yields the same amount of mm-sized dust as the \reference run with a $100M_\oplus$ disk.
Of course, this would solve the mass problem. However, the main argument against this as a solution remains the same: the disk modeled in the \zodisteep run is incompatible with observations. Further, as discussed above, using a steep $\QD$ at sizes larger than about a millimeter lacks justification~-- and, in fact, is at odds with dedicated numerical simulations of collisions at such sizes \citep[e.g.,][and references therein]{benz-asphaug-1999,leinhardt-stewart-2012}.

\section{Conclusions}
\label{Sec:Conclusion}

In this study, we seek to constrain a poorly known tensile strength of dust in debris disks. To this end, we simulate collisional evolution of several fiducial debris disks to see how it responds to changes in the critical fragmentation energy of dust, $\QD$. We arrive at the following conclusions:
\begin{itemize}
\item
We find an upper limit on $\QD$ above which collisional evolution of a disk changes significantly. That limit lies about a order of magnitude above the values typical of standard $\QD$ prescriptions commonly used in debris disk modeling. For a disk with a radius of $100\au$ around a Sun-like star, the threshold lies at $\sim 10^{9...10}\ergg$ for micrometer-sized particles. If debris dust is harder than that, collisions in the outer disk (outside the parent planetesimal ring) are no longer disruptive or cratering. Instead, they are rebounding, i.e., changing the orbits of dust grains while eroding both colliding particles only slightly. This alters the size-spatial distribution of dust tangibly, compared to disks with softer dust.\\
\item 
Collisional evolution of such rebound-dominated disks can be summarized as follows. Collisional cascade in the parent planetesimal ring produces micron-sized dust grains that are sent by radiation pressure into eccentric orbits with pericenters in the birth ring and apocenters farther out, forming a halo of small grains exterior to the main belt. Collisions between the halo grains fragment them into yet smaller dust; radiation pressure leads to size-sorting of these fragments, placing small dust motes in the outer disk in circular orbits. At each distance outside the main belt, disk is composed of grains of a certain size in nearly-circular orbits. The larger the distance, the smaller that size. Both the collisions between the circularized particles themselves and the collisions between the these grains and halo particles are rebounding, making these size-sorted grains in circular orbits long-lived. However, small mass losses in those collisions decrease the sizes of the circularized particles, causing their orbits to slowly expand over time. The circular dust orbits at large stellocentric distances lead to peculiar profiles of dust density (or optical depth, or brightness) very different from those in debris disks where collisions are predominantly disruptive.\\
\item 
In order to see how different mechanical strength of dust would show up in debris disk observations, we analyze typical observables. We find that both long-term decay of the disk brightness with system's age and spectral energy distribution are pretty similar for all $\QD$ models probed in our study.
In contrast, radial brightness profiles of disks turn out to be a sensitive tracer of the dust strength. We show that disks of very hard dust would imply brightness profiles clearly incompatible with those actually observed for resolved debris disks, especially in scattered light.
\end{itemize}

\begin{acknowledgements}
We thank Philippe Thébault and an anonymous reviewer for their thorough, constructive, and speedy reports that greatly helped to improve the paper.     \end{acknowledgements}

\newcommand{\AAp}      {A\&A}
\newcommand{\AApR}     {Astron. Astrophys. Rev.}
\newcommand{\AApS}     {AApS}
\newcommand{\AApSS}    {AApSS}
\newcommand{\AApT}     {Astron. Astrophys. Trans.}
\newcommand{\AdvSR}    {Adv. Space Res.}
\newcommand{\AJ}       {AJ}
\newcommand{\AN}       {Astron. Nachr.}
\newcommand{\AO}       {App. Optics}
\newcommand{\ApJ}      {ApJ}
\newcommand{\ApJL}      {ApJL}%
\newcommand{\ApJS}     {ApJS}
\newcommand{\ApSS}     {Astrophys. Space Sci.}
\newcommand{\ARAA}     {ARA\&A}
\newcommand{\ARevEPS}  {Ann. Rev. Earth Planet. Sci.}
\newcommand{\BAAS}     {BAAS}
\newcommand{\CelMech}  {CMDA}
\newcommand{\EMP}      {Earth, Moon and Planets}
\newcommand{\EPS}      {Earth, Planets and Space}
\newcommand{\GRL}      {Geophys. Res. Lett.}
\newcommand{\JGR}      {J. Geophys. Res.}
\newcommand{\JOSAA}    {J. Opt. Soc. Am. A}
\newcommand{\MaPS}     {Meteoritics \& Plan. Sci.}
\newcommand{\MemSAI}   {Mem. Societa Astronomica Italiana}
\newcommand{\MNRAS}    {MNRAS}
\newcommand{\PASJ}     {PASJ}
\newcommand{\PASP}     {PASP}
\newcommand{\PSJ}      {PSJ}
\newcommand{\PSS}      {Planet. Space Sci.}
\newcommand{\QJRAS}    {Quart. J. Roy. Astron. Soc.}
\newcommand{\RAA}      {Res. Astron. Astrophys.}
\newcommand{\SolPhys}  {Sol. Phys.}
\newcommand{\SolSysRes}{Sol. Sys. Res.}
\newcommand{\SSR}      {Space Sci. Rev.}

\bibliography{ref}
\bibliographystyle{aa}

\begin{appendix}
\nolinenumbers
\section{Mass scaling}
\label{app:mass-scaling}

The collisional evolution, with objects grouped in discrete bins, can be described by a kinetic master equation:
\begin{equation}\label{eq:kinetic-master}
  \frac{\total n_i}{\total t} = \sum_{jk} G_{ijk} n_j n_k - n_i \sum_j L_{ij} n_j,
\end{equation}
where $n$ are the numbers of objects in the bins, with each index $i$ (or $j$ or $k$) being multi-dimensional (masses, pericenter distances, eccentricities, in our case). The constant coefficient $G_{ijk}$ describes how many objects of type $i$ are gained in collisions between $j$ and $k$; $L_{ij}$ describes the losses in collisions between $i$ and~$j$.

Applying a transformation
\begin{equation}\label{eq:mass-scaling}
  n' \equiv xn,
\end{equation}
we obtain
\begin{eqnarray}
  \frac{1}{x}\frac{\total n_i'}{\total t} &=& \sum_{jk} G_{ijk} \frac{n_j' n_k'}{x^2} - \frac{n_i'}{x} \sum_j L_{ij} \frac{n_j'}{x}\nonumber\\
  x\frac{\total n_i'}{\total t} &=& \sum_{jk} G_{ijk} n_j' n_k' - n_i' \sum_j L_{ij} n_j',
\end{eqnarray}
which can be made equivalent to eq.~(\ref{eq:kinetic-master}) if we let
\begin{equation}\label{eq:time-scaling}
  t' \equiv t/x.
\end{equation}
This is the scaling law~(\ref{shiftLaw}) used by \citet{loehne-et-al-2007} and \citet{krivov-et-al-2008}. It holds if all gains and losses are caused by two-body collisions. Where other effects are important for the evolution, such as Poynting--Robertson drag or radiation pressure blowout, the scaling law no longer holds. 

For unbound grains, now denoted by the index $\beta$, losses are dominated by dynamical ejection from the system, which occurs on a (typically short) timescale $T_\beta$:
\begin{equation}
  \frac{\total n_\beta}{\total t} = \sum_{jk} G_{\beta jk} n_j n_k - \frac{n_\beta}{T_\beta},
\end{equation}
where we assume $j,k\neq \beta$, thus neglecting contributions of the $\beta$ grains to their own gains.
In equilibrium, where $\dot{n_\beta} = 0$, we have
\begin{equation}
  n_\beta(t) = T_\beta \sum_{jk} G_{\beta jk} n_j n_k,
\end{equation}
such that the transformations~(\ref{eq:mass-scaling}) and (\ref{eq:time-scaling}) lead to 
\begin{equation}
  n_\beta'(t') = T_\beta \sum_{jk} G_{\beta jk} n_j' n_k' = x^2 n_\beta(t).
\end{equation}
This shows that the number of unbound grains scales with $x^2$ instead of $x$, justifying eq.~(\ref{shiftLawBeta}).

\section{The role of rebounding collisions}
\label{sec: app rebounding}

As found in Section~\ref{spatialDistribution}, the \zodisteep run tends to form circular orbits of small dust in the outer disk. A natural guess is that, since the dust in that run has a high $\QD$, disruptive collisions may not be efficient, and other collisional outcomes may take over.
To investigate this, an additional \zodisteep run was performed with rebounding collisions turned off. 

\begin{figure}[h!]
    \centering
    \includegraphics[width=0.99\linewidth]{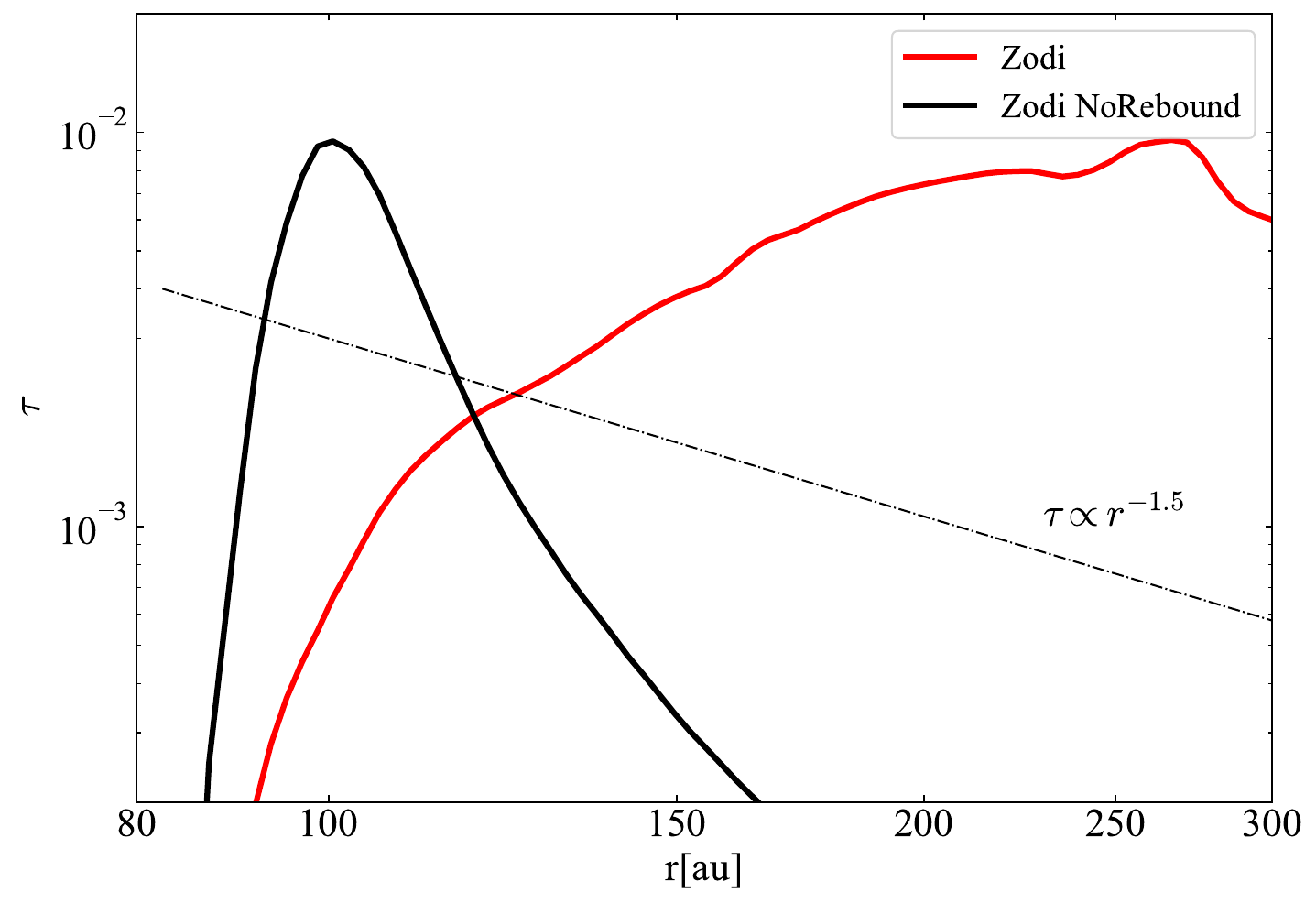}
    \caption{
    Radial profiles of normal optical depth in the \zodisteep run (red lines) and an additional \zodisteep run without rebounding collisions (black). An idealized profile (eq.~\ref{tau1.5}) is shown for comparison.
    }
    \label{fig:tauNoRebounding}
\end{figure}

\begin{figure}[h!]
\centering
    \includegraphics[width=0.99\linewidth]{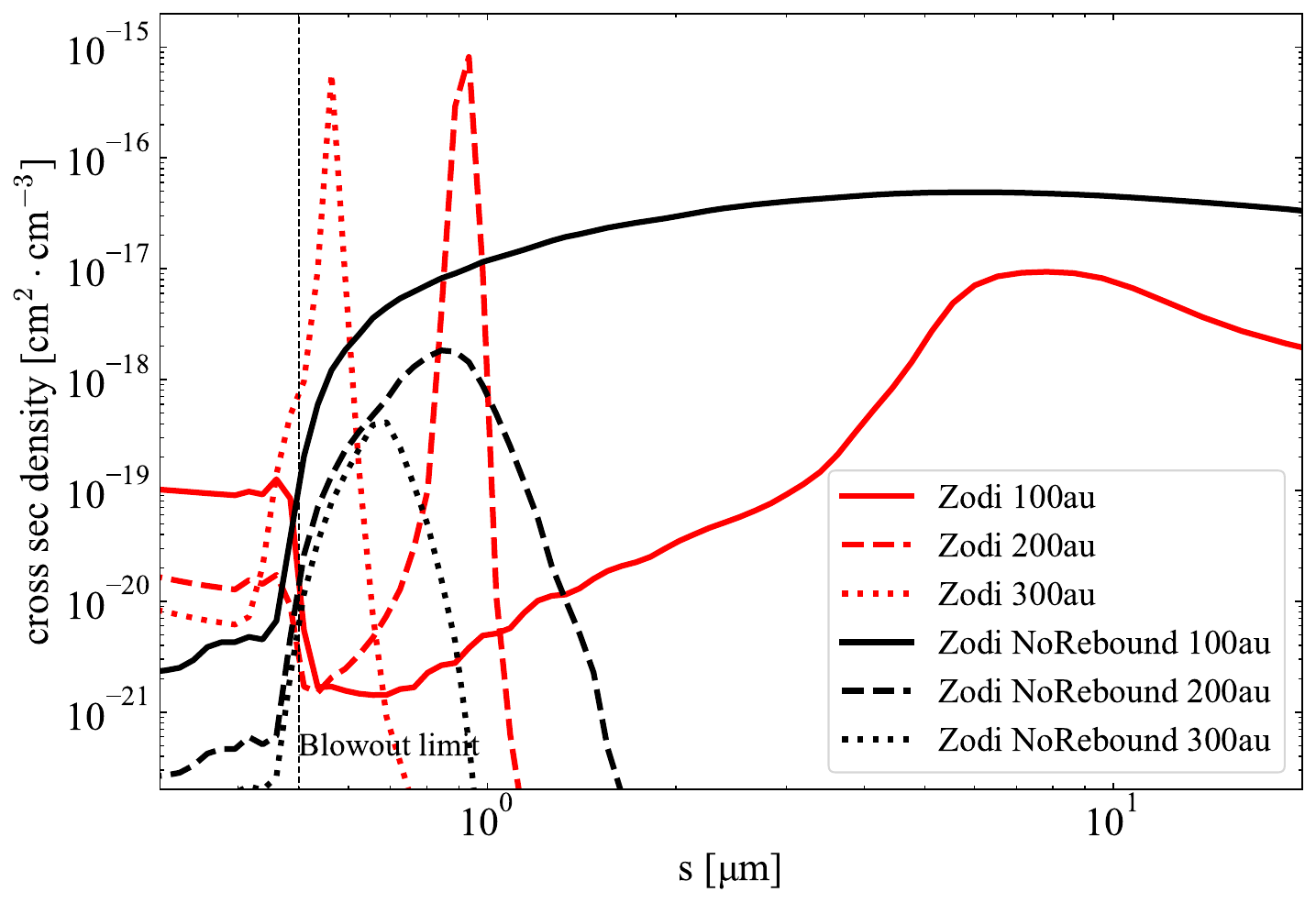}
    \caption{
    Size distribution of small grains ($s<10\mum$) in the \zodisteep run (red lines) and an additional \zodisteep run without rebounding collisions (black) at three selected distances: $100\au$ (solid lines), $200\au$ (dashed), and $300\au$ (dotted).
    }
    \label{fig:CrossSecNoRebound}
\end{figure}

\begin{figure}[h!]
\centering
    \includegraphics[width=0.99\linewidth]{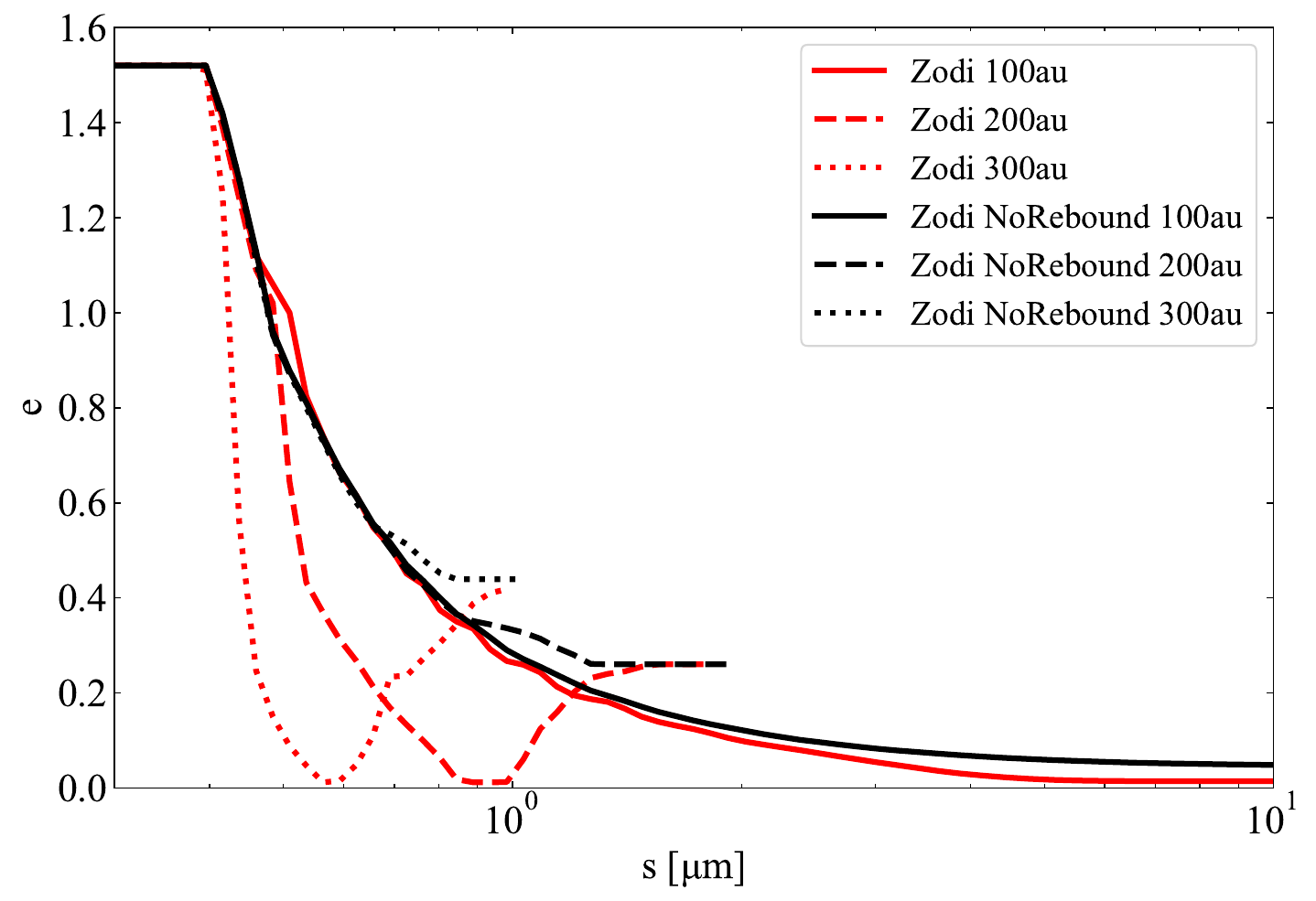}
    \caption{
    Average eccentricity in the \zodisteep run with rebounding collisions (red lines, same as in Fig.~\ref{fig:CrossSec200au_Ecc}) and without them (black lines) plotted against particle size at different distances.
    }
    \label{fig:EccNoRebounding}
\end{figure}

\clearpage
Figure~\ref{fig:tauNoRebounding} compares the optical depth profiles for the \zodisteep run with and without rebounding collisions. Without rebounding collisions, the optical depth profile exhibits the ``normal'' decrease with stellocentric distance.
Figure~\ref{fig:CrossSecNoRebound} shows the size distribution in the two runs. The most visible difference is that without rebounding collisions, the size distributions become smooth and similar to those in the \reference and \fomaref runs (cf. Fig.~\ref{fig:CrossSec200au_Ecc} left).
In Fig.~\ref{fig:EccNoRebounding}, the mean eccentricities of different-sized particles with and without rebounding collisions are plotted, at several distances. The peculiar behavior described in Section~\ref{spatialDistribution} disappears when rebounding collisions are switched off.
These results demonstrate that rebounding collisions, as opposed to disruptive ones, are the dominant collisional outcome in the \zodisteep run.

\section{Circular orbits in the outer disk}
\label{sec: app circular rings}

\begin{figure*}[htp!]
    \centering
    \includegraphics[width=0.8\linewidth]{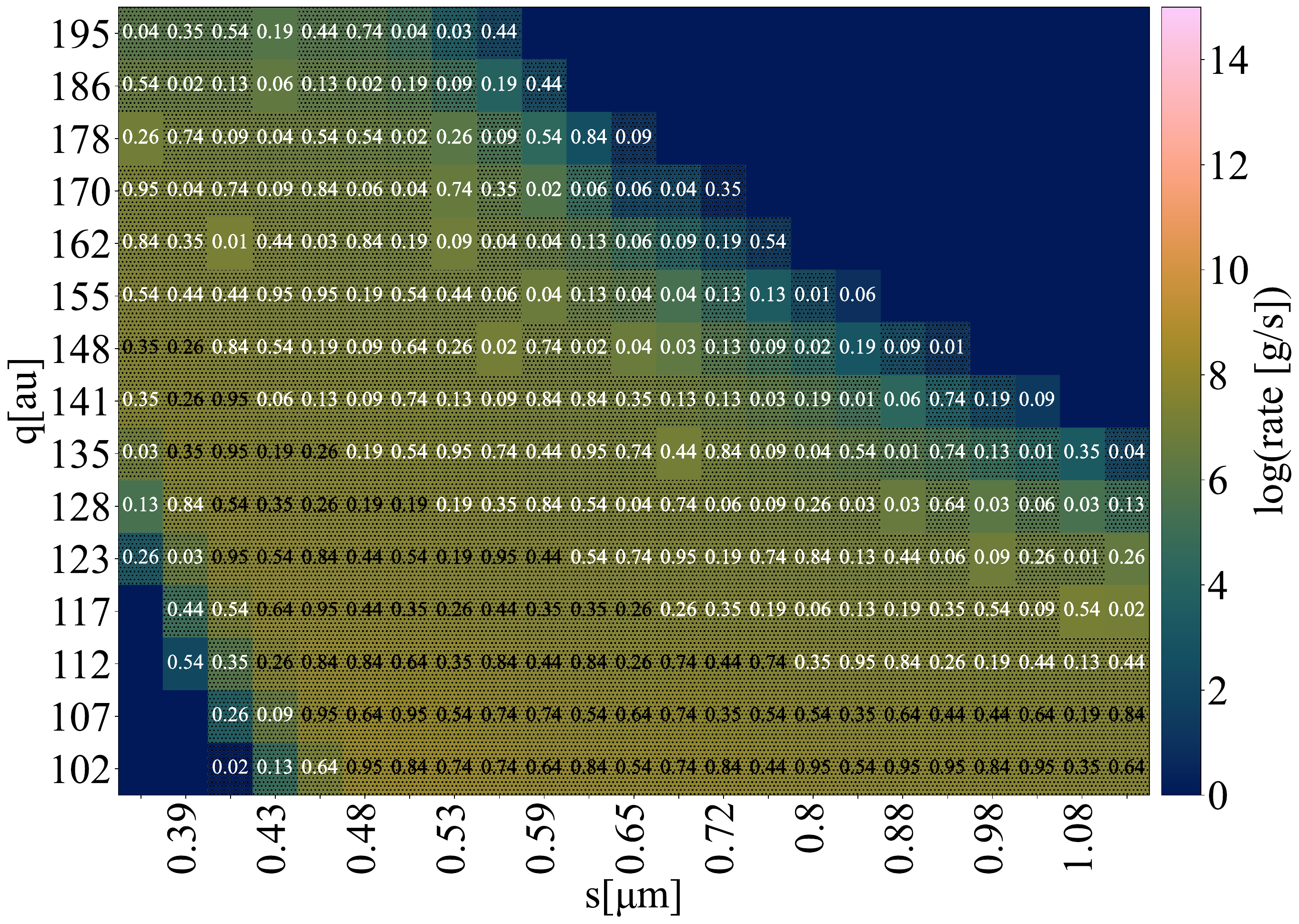}
    \includegraphics[width=0.8\linewidth]{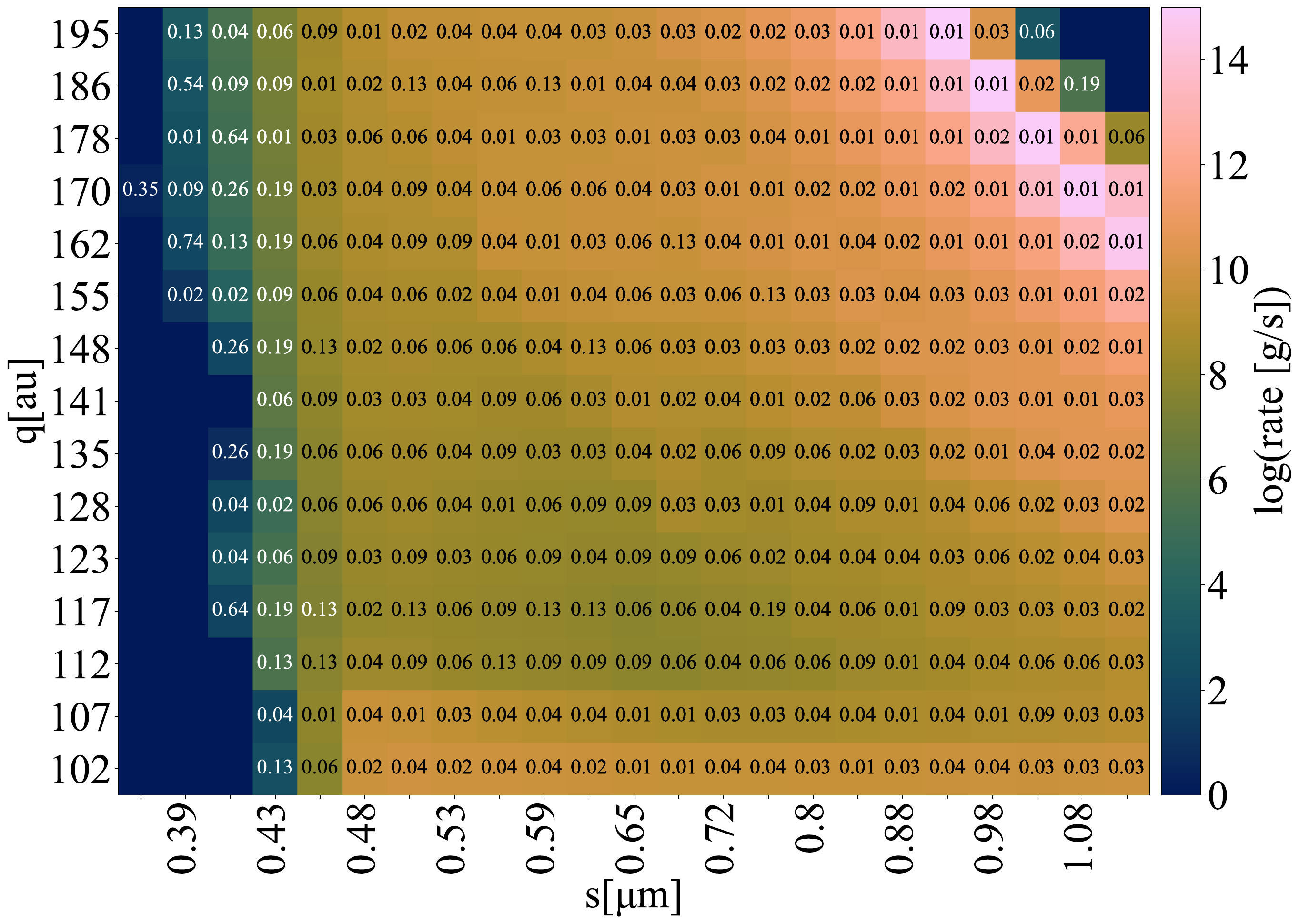}
    \caption{
    Collisions that remove particles from the $(s=0.93\mum, e=0.35, q=102\au)$ bin at $T=100\Myr$ in the \reference (left) and \zodisteep runs (right). 
    The vertical and horizontal axes are the pericenter distance $q$ and size $s$ of the grains that collide with grains in that bin, respectively. Colors represent the collisional rates. These rates are summed over all collision partner's eccentricities, and the numbers in the bins represent the eccentricity of the projectile with the highest collision rate. Bins where the collision with the projectile with the highest collision rate is disruptive or cratering are hatched while the ones where it is rebounding are not.
    }
    \label{fig:LossPlots}
\end{figure*}

Here we investigate in more detail the collisions in the outer disk of the \reference and \zodisteep runs, in order to get more insights into the absence of circular orbits in the former one and their presence in the latter.

We first remind the reader how ACE simulations work. All disk particles are distributed into a three-dimensional grid of bins $(m,q,e)$, where $m$ is the mass of a particle, $q$ the pericenter distance of its orbit, and $e$ the orbital eccentricity.
For every pair of bins, collisions with four collisional outcomes can occur: disruptive, cratering, rebounding, and sticking (eqs.~\ref{eq:disrupt}--\ref{eq:stick}), of which only the former three are relevant for our simulations. When a collision happens, the two colliders are removed from their bins and the appropriate bins are incremented to accommodate the particles created in that collision.
Each of those collision types can happen a certain number of times during one time step, which associates each collision type with a certain collision rate.

As an example, we consider
$0.9\mum$ particles with pericenters at $100\au$ and apocenters at $200\au$, giving us an eccentricity of $e=1/3$.
(Note that, due to the discrete grid of the simulation, we actually consider the bin $s=0.93\mum, e=0.35, q=102\au$.)
This bin was chosen because in the \zodisteep run $0.9\mum$-sized particles get in circular orbits of radius $200\au$ (see Fig.~\ref{fig:CrossSec200au_Ecc}). We wish to see further evidence for the existence of these orbits in the \zodisteep run, as well as the absence thereof in the \reference one.
Collisions experienced by the grains in that bin are visualized in
Fig.~\ref{fig:LossPlots} as a heatmap\footnote{To minimize visual distortion, heatmaps in this study employ the perceptually uniform colormap batlow \citep{crameri_2023_8409685}.}. It shows how often
collisions in the bin occur with projectiles of different sizes and pericenter distances, and what the typical eccentricities of those impactors are.

We start by discussing the \reference run (Fig.~\ref{fig:LossPlots} left).
Inspecting how the bins are incremented and decremented, we found that the $(s=0.93\mum, e=0.35, q=102\au)$ ring experiences all relevant collisional outcomes~-- destructive, cratering, and rebounding.
This means that in the \reference run, collisions efficiently reduce particle mass, creating fragments that are too small to get injected by radiation pressure in circular orbits.
Interestingly, the map demonstrates that the halo particles in the \reference run get destroyed by highly eccentric particles with various sizes and various pericenter distances~-- predominantly by $\sim 0.5\mum$-sized particles with pericenters in the main belt and very high eccentricities.

We now consider the \zodisteep run. Our analysis of the bin evolution confirmed that only rebounding collisions are at work, as dust in the bin in question is too hard to undergo destructive and cratering collisions. Rebounding collisions gradually erode the grains, slightly reducing their sizes. In particular, they produce the particles with $\beta = \beta^\prime$ (eq.~\ref{betaprime}), which get in circular orbits.
The heatmap in Fig.~\ref{fig:LossPlots} (right) also shows that the vast majority of collisions experienced by the halo particles in the $(s=0.93\mum, e=0.35, q=102\au)$ bin~-- i.e.\ those that act as a source for the circular orbits $200\au$ in radius~-- are particles that have different sizes, but all move in nearly circular orbits of different radii in the outer disk.

In addition to the already mentioned differences between the two heatmaps, it is clearly visible that there are many more collisions happening in the \zodisteep run than in the \reference one. This is because there are more small dust particles that act as possible collision partners in the \zodisteep run, in particular near the apocenter at $200\au$ where the collisional rates are highest.

\end{appendix}
\end{document}